\documentclass{siamltex}
\usepackage{amsmath}
\usepackage{amssymb}
\usepackage{amsfonts}
\usepackage{graphicx}
\usepackage{color}

\newcommand{\comment}[1]{} % makes its argument disappear

\def\bq{\begin{quotation}}
\def\eq{\end{quotation}}

\definecolor{blue}{rgb}{0,0,1}
\definecolor{pink}{rgb}{0.9,0.5,0.5}

\def\a{\alpha}

\def\z{\zeta}

\def\k{\kappa}

\def\s{\sigma}

\def\fs{\phi}

\def\Y{\Psi}

\def\2nm#1{\|#1\|_2}

\newcommand{\ars}[1]{\left[ \begin{array}{#1}}
\newcommand{\are}{\end{array} \right] }
\newcommand{\oars}[1]{\begin{array}{#1}}
\newcommand{\oare}{\end{array}}

\newcommand{\eqs}{\begin{eqnarray}}
\newcommand{\eqe}{\end{eqnarray}}
\newcommand{\eqsn}{\begin{eqnarray*}}
\newcommand{\eqen}{\end{eqnarray*}}

\newcommand{\ens}{\begin{enumerate}}
\newcommand{\ene}{\end{enumerate}}

\newcommand{\its}{\begin{itemize}}
\newcommand{\ite}{\end{itemize}}

\newcommand{\des}{\begin{description}}
\newcommand{\dee}{\end{description}}

\makeatletter
\newif\if@borderstar
\def\bordermatrix{\@ifnextchar*{%
  \@borderstartrue\@bordermatrix@i}{\@borderstarfalse\@bordermatrix@i*}%
}
\def\@bordermatrix@i*{\@ifnextchar[{%
  \@bordermatrix@ii}{\@bordermatrix@ii[()]}
}
\def\@bordermatrix@ii[#1]#2{%
  \begingroup
    \m@th\@tempdima8.75\p@\setbox\z@\vbox{%
      \def\cr{\crcr\noalign{\kern 2\p@\global\let\cr\endline }}%
      \ialign {$##$\hfil\kern 2\p@\kern\@tempdima & \thinspace %
      \hfil $##$\hfil && \quad\hfil $##$\hfil\crcr\omit\strut %
      \hfil\crcr\noalign{\kern -\baselineskip}#2\crcr\omit %
      \strut\cr}}%
    \setbox\tw@\vbox{\unvcopy\z@\global\setbox\@ne\lastbox}%
    \setbox\tw@\hbox{\unhbox\@ne\unskip\global\setbox\@ne\lastbox}%
    \setbox\tw@\hbox{%
      $\kern\wd\@ne\kern -\@tempdima\left\@firstoftwo#1%
        \if@borderstar\kern2pt\else\kern -\wd\@ne\fi%
      \global\setbox\@ne\vbox{\box\@ne\if@borderstar\else\kern 2\p@\fi}%
      \vcenter{\if@borderstar\else\kern -\ht\@ne\fi%
        \unvbox\z@\kern-\if@borderstar2\fi\baselineskip}%
        \if@borderstar\kern-2\@tempdima\kern2\p@\else\,\fi\right\@secondoftwo#1 $%
    }\null \;\vbox{\kern\ht\@ne\box\tw@}%
  \endgroup
}
\makeatother
\title{Improved Scaling for Quantum Monte Carlo on Insulators
\thanks{This material is based
upon work supported by the National Science Foundation under
Grant No. NSF-EAR 0530643
and by the Materials Computation Center at the University of Illinois}}

\author{
Kapil Ahuja\footnotemark[2] \and Bryan K. Clark\footnotemark[3]
\and Eric de Sturler\footnotemark[4] \and David M.
Ceperley\footnotemark[5]
\and Jeongnim Kim\footnotemark[6]}  %DMC

\begin{document}
\maketitle

\renewcommand{\thefootnote}{\fnsymbol{footnote}}
\footnotetext[2]{Department of Mathematics, Virginia Tech,
Blacksburg, VA 24061 (kahuja@vt.edu).} %
\footnotetext[3]{Department of Physics, University of Illinois
at Urbana-Champaign, Urbana, IL 61801 and Physics Department,
Princeton University, Princeton, NJ 08544
(bclark@princeton.edu).} %
\footnotetext[4]{Department of Mathematics, Virginia Tech,
Blacksburg, VA 24061 (sturler@vt.edu).} %
\footnotetext[5]{Department of Physics, University of Illinois
at Urbana-Champaign, Urbana, IL 61801 (ceperley@uiuc.edu).} %
\footnotetext[6]{National Center for Supercomputing Applications,
Urbana IL 61801 (jnkim@uiuc.edu).} %DMC
\renewcommand{\thefootnote}{\arabic{footnote}}

\begin{abstract}
Quantum Monte Carlo (QMC) methods are often used to calculate
properties of many body quantum systems. The main cost of many
QMC methods, for example the variational Monte Carlo (VMC)
method, is in constructing a sequence of Slater matrices and
computing the ratios of determinants for successive Slater
matrices. Recent
%ly, there has been some
%discussion about
work has improved the scaling of constructing Slater matrices
for insulators so that the cost of constructing Slater matrices
in these systems is now linear in the number of particles,
whereas computing determinant ratios remains cubic in the
number of particles.
%albeit with a much smaller constant.
With the long term aim of simulating much larger systems, we
improve the scaling of computing the determinant ratios in the
VMC method for simulating insulators by using preconditioned
iterative solvers.

The main contribution of this paper is the development of a
method to {\em efficiently} compute for the Slater matrices a
sequence of preconditioners that make the iterative solver {\em
converge rapidly}. This involves cheap preconditioner updates,
an effective reordering strategy, and a cheap method to monitor
instability of ILUTP preconditioners. Using the resulting
preconditioned iterative solvers to compute determinant ratios
of consecutive Slater matrices reduces the scaling of QMC
algorithms from $O(n^3)$ per sweep to roughly $O(n^2)$, where
$n$ is the number of particles, and a sweep is a sequence of
$n$ steps, each attempting to move a distinct particle. We
demonstrate experimentally that we can achieve the improved
scaling without increasing statistical errors. Our results show
that preconditioned iterative solvers can dramatically reduce
the cost of VMC for large(r) systems.

\end{abstract}


% 65C05 Monte Carlo Methods
% 81Q05 Closed and approximate solutions to the Schr¨odinger, Dirac, Klein-Gordon and other quantum-mechanical equations
% 65F40 Determinants
% 65F50 Sparse Matrices
% 82B80 Monte Carlo
% 11K45 Monte Carlo methods

\pagestyle{myheadings} \thispagestyle{plain} \markboth{Ahuja,
Clark, de Sturler, Ceperley, and Kim}{Efficient Variational
Monte Carlo}

\section{Introduction}\label{sec:intro}
Quantum Monte Carlo (QMC) methods \cite{thijssen:book,Ha94},
like the Variational Monte Carlo method (VMC), produce highly
accurate quantitative results for many body systems. They have
a wide range of applications, including the study of the
electronic structure of helium, molecules, solids, and lattice
models. The QMC method is computationally expensive, and
current algorithms and computers limit typical system sizes for
fermionic systems to about a thousand particles
\cite{foulkes:review,mitas:few,scalettar:few}.
%The
%lack of sufficient computing power is the main constraint in
%simulating larger systems, such as, for example, computing the
%ground state energy of a solid using the variational Monte
%Carlo method.
There are two main bottlenecks to simulating larger systems.
The first is constructing a sequence of millions of so-called
Slater matrices, and the second is computing the ratios of
determinants of successive Slater matrices. We will briefly
describe VMC, Slater matrices, and other relevant concepts in
section~\ref{sec:vmc}, but first we discuss the scaling of
these bottlenecks. Our longer term aim is to develop methods
for the efficient simulation of systems with $10^4$ to $10^5$
particles (on high-end computers).

Let $n$, the system size, be the number of particles
(electrons) in the system, which also equals the number of
orbitals. Each orbital is a single particle wave function. For
simplicity we ignore spin; incorporation of spin is
straightforward. In VMC, we estimate observables by
conditionally moving particles one by one and accumulating
`snapshots' of the observables; see section~\ref{sec:vmc}.
We define the attempt to move all the particles in the system once as a sweep.
% Mention briefly the idea of MC, generating moves and accepting or rejecting them
% provide some references for MC
The cost of constructing Slater matrices depends on the type of
basis functions used for building the orbitals. We discuss this
dependency in the next section; however, the cost for the
generic case is $O(n^2)$ per sweep \cite{williamson:linear}
(constant cost per element in the matrix). Recent physics
papers \cite{dario:linear, dario:splines, williamson:linear}
discuss methods to reduce this cost to $O(n)$ by optimizing the
orbitals, such that the matrix is optimally sparse. Therefore
only a linear number of elements in the matrix must be filled
and computing the matrix becomes cheap. This can only be shown
to be rigorous for certain physical systems (like insulators).
These methods are referred to as `linear scaling' or `O(n)'
methods, because, in many cases,
% DMC for sufficiently small system sizes,
the cost of computing the Slater matrices dominates. However,
these methods are not truly linear, since the cost of computing
the determinant ratios is still $O(n^3)$, which will dominate
the cost of the VMC method for larger $n$. In this paper, we
focus on reducing this cost for insulators (or, more generally,
for systems with sparse Slater matrices) to $O(n^2)$.

The VMC method generates a sequence of matrices,
\begin{eqnarray}\label{eq:MatrixUpdate_1}
    A_{k+1} & = & A_{k} + e_{i_{k}} u_{k}^T ,
\end{eqnarray}
where $k$ indicates the Monte Carlo step or particle move,
$A_k$ and $A_{k+1}$ are the Slater matrices before and after
the proposed move of particle $i_{k}$, $e_{i_{k}}$ is the
corresponding Cartesian basis vector, and $u_{k}$
%\begin{equation}\label{matrixUpdateU}
%    u_{k}(j) = \phi_j(r'_{i_{k}}) - \phi_j(r_{i_{k}})
%\end{equation}
gives the change in row $i_k$ resulting from moving particle
$i_k$. The acceptance probability of the (conditional) move
depends on the squared absolute value of the determinant ratio
of the two matrices,
\begin{equation}\label{eq:DetRatio}
    \frac{| \det A_{k+1} |^2}{|\det A_{k}|^2} = \left| 1 +  u_{k}^T A_{k}^{-1}
    e_{i_{k}}\right|^2 .
\end{equation}
The {\em standard algorithm} in VMC \cite{CepChes77} uses the explicit
inverse of the Slater matrix, $A_k^{-1}$, to compute this ratio, and it updates
this inverse according to the Sherman-Morrison formula \cite[p.
50]{golub:book} if the particle move is accepted, resulting in $O(n^3)$ work per sweep.
For stability, the inverse is occasionally recomputed from scratch
(but such that it does not impact the overall scaling). For
systems with fewer than a thousand particles this method
is practical.
Recently, in \cite{nukula:slater} a variant of this
algorithm was proposed that accumulates the multiplicative
updates to the exact inverse,
\[
  A_{k+1}^{-1} = (I - \frac{1}{1 + u_k^TA_k^{-1}e_{i_k}}
  A_k^{-1}e_{i_k}u_k^T) A_k^{-1} ,
\]
and applies those recursively to compute the ratio
(\ref{eq:DetRatio}), an approach well-known in numerical
optimization for Broyden-type methods \cite[p.
88]{kelley:book}. The same idea is also discussed in
\cite[Appendix A]{BaiSca09}. This approach requires $O(kn)$ work
for the $k^{th}$ Monte Carlo step, and therefore $O(k^2 n)$ work for the
first $k$ Monte Carlo steps. Hence, a single sweep ($n$
Monte Carlo steps) takes $O(n^3)$ work. If the inverse is not recomputed
once per sweep, the approach will actually scale worse than the
standard algorithm. So, this method does not decrease
the total number of required operations (or scaling), but often
has superior cache performance resulting in an increase in
speed.

%So, even after a single
%sweep, this algorithm becomes more expensive than the standard
%algorithm, and hence is not competitive for VMC simulations
%where hundreds of sweeps have to be performed.
% \footnote{The algorithm of~\cite{nukula:slater} mainly aims
% at diffusion Monte Carlo (DMC) simulations with relatively
% few particle updates.
% Also in \cite{BaiSca09} the number of steps is small
% compared with the total system size.}
In this paper, we propose an algorithm for computing
determinant ratios for insulating systems in the VMC method
with roughly $O(n^2)$ work per sweep, providing a significant
improvement in the order of complexity over the standard
algorithm; see the discussion of complexity and
tables~\ref{table:cost} and \ref{table:time} in
section~\ref{sec:expmnts}.

Rather than keeping and updating the inverse of the Slater
matrices, we can compute (\ref{eq:DetRatio}) in step $k$ by
solving $A_k z_k = e_{i_k}$ and taking the inner product of
$z_k$ with $u_k$. Since, for certain cases (like insulators),
the Slater matrices with optimized orbitals are sparse, and the
relative sparsity increases with the number of particles and
nuclei in a system for a given material, preconditioned
iterative methods are advantageous, especially since the
accuracy of the linear solve can be modest. As we will show,
convergence is rapid with a good preconditioner, and hence we
use preconditioned full GMRES \cite{SaadSchu86}. However, other
Krylov methods are equally possible and might be more
appropriate for harder problems. {\em The main challenge in our
approach is to generate, at low cost, a sequence of
preconditioners corresponding to the Slater matrices that
result in the efficient iterative solution of each system}.
%It turns out that the sequence of particle moves
%leads to matrices for which incomplete decompositions are of
%poor quality.
We explain how to resolve this problem in
section~\ref{sec:AlgoImprove}.

In section~\ref{sec:vmc}, we briefly discuss the VMC method. We
discuss the efficient computation of a sequence of
preconditioners and their use in computing determinant ratios
by iterative solves in section~\ref{sec:AlgoImprove}.
Section~\ref{sec:expmnts} gives numerical results that
demonstrate experimentally the efficiency and favorable scaling
of our approach. We also show that our approach works in
practice.
% does not lead to an increase of statistical error.
Finally, we discuss
conclusions and future work in section \ref{sec:conclusion}.

\section{The Variational Monte Carlo Method}\label{sec:vmc}
The VMC method for computing the ground-state expectation value
of the energy combines variational optimization with the Monte
Carlo algorithm to evaluate the energy and possibly other
observables \cite{foulkes:review,thijssen:book}.  VMC is one of
many forms of quantum Monte Carlo.  Although our focus will be
on variational Monte Carlo, our results naturally transfer to
diffusion Monte Carlo methods \cite{foulkes:review} as well.

The inner loop of VMC involves sampling over configurations $R$
with  the probability density $|\Y_\a(R)|^2$ induced by the
many body trial wave function $\Psi_\alpha(R)$
\cite{thijssen:book}. Here, $\alpha$ denotes the vector of
variational parameters over which we minimize, and $R$ is a
$3n$-dimensional vector representing the coordinates of all the
particles in the system. This sampling is done using a Markov
Chain Monte Carlo approach.  For each independent configuration
$R$ generated from the Markov Chain, observables ${\cal
O}(R,\Psi_\alpha)$ are computed and averaged.  Although there
are many possible observables, our discussion will focus on the
the local energy, $E_{L,\alpha}$.  The average of
$E_{L,\alpha}$ is the function being optimized in the outer
loop and consequently essential for any VMC calculation. In the
rest of this section, we will further detail the separate
pieces of this process with a particular focus on the
computational bottlenecks.
%
%In this section, we will discuss the nature of the many body wave-function
%$\Psi_\alpha(R)$ and the local energy $E_l$ in addition to seeing how ratios
%of these wave functions arise in VMC.
%
\subsection{Many Body Wave Function}

Although wave functions come in a
variety of forms, the most common form is
\begin{eqnarray}\label{eq:WaveFunction}
    \Psi_\a(r_1,r_2,...,r_n) & = & \exp\left(\sum_{ij}f(r_i-r_j)\right)\det(A),
    \quad\mbox{ with  }
\\ \label{eq:SlaterMatrix}
  A & = & \left(
            \begin{array}{cccc}
              \fs_1(r_1) & \fs_2(r_1) & \ldots & \fs_n(r_1) \\
              \fs_1(r_2) & \fs_2(r_2) & \ldots & \fs_n(r_2) \\
              \vdots     &            & \ddots & \vdots \\
              \fs_1(r_n) & \fs_2(r_n) & \ldots & \fs_n(r_n)
            \end{array}
          \right) ,
\end{eqnarray}
where $f$ is called the Jastrow factor \cite{CepChes77}\cite[p.
319]{thijssen:book}, $r_i$ represents the coordinates of
particle $i$ (part of the vector $R$), $A$ is called a Slater
matrix (with elements $a_{i,j} = \fs_j(r_i)$), and its
determinant is called a Slater determinant. Each function
$\fs_j$ is a scalar function of position %inside the periodic box of the simulation
and is referred to as
a single particle orbital or single particle wave function.
% EdS rephrase the following
The calculation optimizes over these single particle orbitals
as functions of the parameter vector $\a$. In this paper, we
assume that the particles are confined to a cube and
periodic boundary conditions are used.  %DMC

%The choice of basis functions for the single-particle orbitals
%plays an important role in forming the trial wave
%function~\cite{foulkes:review}. Two popular options for the
%basis of single-particle orbitals are the plane wave basis and
%the Gaussian basis~\cite{williamson:linear}. The plane wave
%basis is comprised of functions of the form
%\begin{equation}\label{planeWave}
%    %\exp{(i G_j \cdot r)},
%    e^{i G_j \cdot r} ,
%\end{equation}
%where $r$ is spatial position and the $G_j$ are the so-called
%reciprocal lattice vectors \cite{foulkes:review}. These are the
%wave vectors for Fourier expansions of functions that are
%periodic on the primitive crystal lattice of the solid being
%analyzed (the periodicity of its nuclei positions)
%\cite{kittel:book,thijssen:book}.

% \el
The nature of the single particle orbitals depends on the
physical system being simulated. (Since the determinants of the
Slater matrices are invariant, up to the sign, under unitary
transformations, we can only talk about properties of the set
of orbitals.) Two important general classes of  electronic
systems are metals and insulators.  In a metal, the single
particle orbitals extend throughout all of space;  there is no
unitary transformation such that each orbital will have compact
support.  Moreover, forcing compact support by maximally
localizing the orbitals and truncating them outside a fixed
sphere can induce serious qualitative errors, e.g., changing
the system from a metal to an insulator. In contrast, for an
insulator there exists a choice of single particle orbitals
that has compact support over small regions of space (or
truncating them to introduce such local support produces
minimal errors). As the size of the system is made large, the
spread of the individual orbitals
remains fixed, which leads to increasingly sparse matrices. %DMC

Previous work
\cite{dario:splines,dario:linear,williamson:linear} has
developed methods to find unitary transformations that minimize
the spread of the single particle orbitals. These methods often
go by the name {\em linear scaling methods}, although in
practice they simply minimize the number of non-zero elements
in the Slater matrix. Although this makes computing and
updating the Slater matrix cheaper, the determinant computation
still scales as $O(n^3)$. Nonetheless, this is an important
advance, which means that, for an insulating system, the Slater
matrix can be made sparse. In this work, we focus on insulators
and we leverage this starting point in our work with the
assumption that our single particle orbitals are maximally
localized.

The specific details of the single particle wave functions
depend sensitively on the exact material being simulated and
are unimportant for our study. Instead, we use a set of
Gaussians centered on $n$ points tiled in space on a b.c.c.
(body centered cubic) lattice as representative insulator
single particle orbitals:
\begin{equation}\label{gaussian}
   \fs_j(\textbf{r}) =  e^{-k \| \textbf{r}-\textbf{Z}_j \|^2} ,
   \quad \mbox{for } j = 1 \ldots n,
\end{equation} %DMC
%where $d_{ij}$ is the distance between particle $i$ and the
where the $\textbf{Z}_j$ are the lattice positions (i.e.,
physically the position of nucleus $j$), and $k$ determines the
rate of decay of $\fs_j$. We will truncate these Gaussian
functions so that they vanish outside some fixed radius (see
section~\ref{sec:expmnts}). These single particle orbitals have
the same qualitative features as realistic single particle
orbitals for insulators, have nice analytical properties that
allow for easier analysis, and a solution to this problem will
almost certainly translate to generic insulators. These
Gaussians have a tuning parameter $k$ that determines their
(effective) width and physically the bandwidth of the
insulator. For these studies, we focus on $k=1$. (We remind the
reader that we do not take spin degrees of freedom into
account; in realistic simulations, one would have separate
determinants for spin up and spin down electrons.) Furthermore,
we choose the unit of length such that the density of electrons
is $3/(4\pi)$. This implies that the b.c.c. lattice spacing is
$2.031$ and the nearest neighbor distance between
lattice/orbital positions is $1.759$. In this case, the Slater matrices
depend on three choices (parameters), first, the type of lattice defining the
orbital/nuclei positions, second, the spacing of the lattice positions, and, third,
the decay rate of the Gaussian orbitals defined by $k$. After fixing the type
of lattice (body centered cubic), physically, only the
spacing of the orbital centers relative to the decay of the
Gaussians is relevant. Hence, for analyzing the dependence of matrix
properties on the parameters below, we need only vary $k$.

% **NOTE about units.  We choose units of length so that the density
% of electrons is 0.238=$3/(4\pi)$. This implies
% that the b.c.c. lattice spacing is 2.031 and the nearest neighbor distance is 1.759.

\begin{figure}[hbt]
    \centering
    \includegraphics[height=2.5in,width=2.5in]{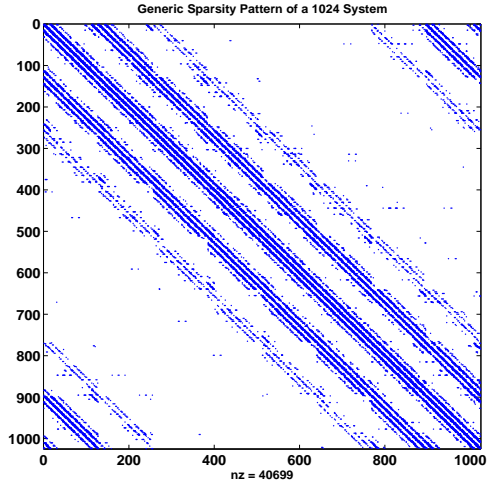}
    \caption{Sparsity pattern for a typical configuration of our system for 1024 particles with $k=1$.
    For smaller $k$ the pattern remains more or less the same, while the number of nonzeros in the matrix
    increases. For larger $k$ the pattern also remains more or less the same, while the number of
    nonzeros in the matrix decreases.}
    \label{figure:sparsity}
\end{figure}
\begin{figure}[hbt]
    \centering
    \includegraphics[height=3in,width=2in,angle=-90]{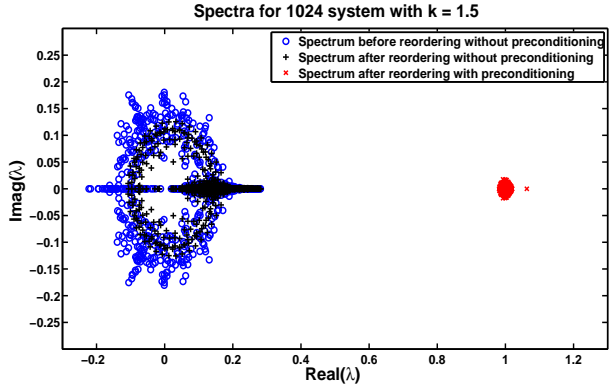}
    \caption{Spectra for a typical configuration of our system for 1024 particles with $k=1.5$,
    before and after reordering, and with preconditioning for the reordered system.}
    \label{figure:spec_k=15}
\end{figure}
\begin{figure}[bt]
    \centering
    \includegraphics[scale=0.30]{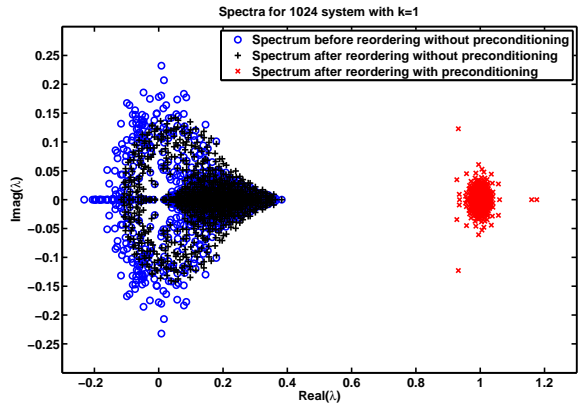}
    \caption{Spectra for a typical configuration of our system for 1024 particles with $k=1$,
    before and after reordering, and with preconditioning for the reordered system.
    The value $k=1$ is used in the experiments reported in section~\ref{sec:expmnts}.}
    \label{figure:spec_k=1}
\end{figure}
\begin{figure}[bth]
    \centering
    \includegraphics[height=3in,width=2in,angle=-90]{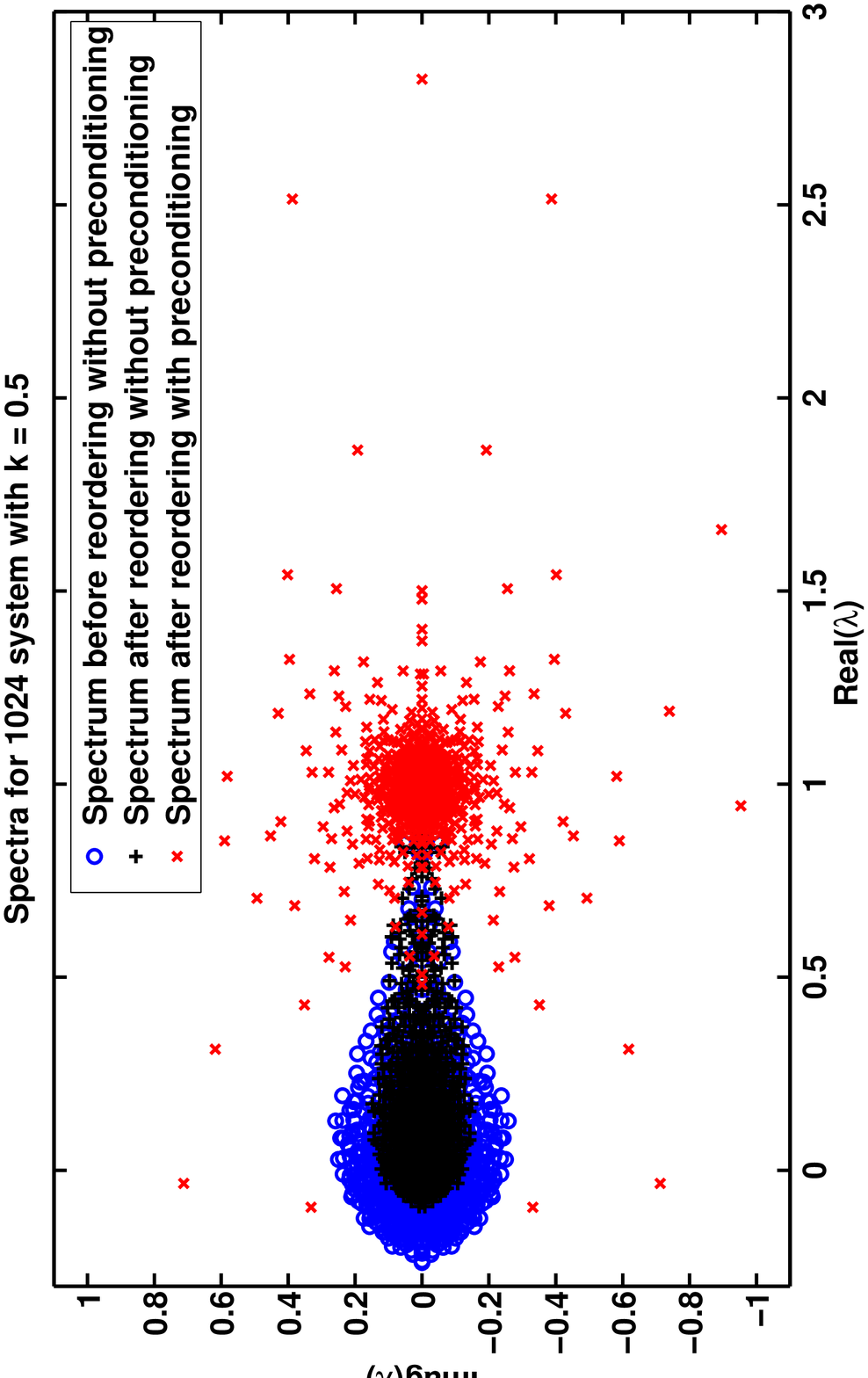}
    \caption{Spectra for a typical configuration of our system for 1024 particles with $k=0.5$,
    before and after reordering, and with preconditioning for the reordered system.
    Note that the scale of this picture differs from the previous two. In addition,
    a lonely eigenvalue of the preconditioned system after reordering at $11.77$ has been
    left out to obtain a better scale.}
    \label{figure:spec_k=05}
\end{figure}

Below, we provide a brief overview of properties of Slater matrices arising from the
choices discussed above. First, in figure~\ref{figure:sparsity},
we show the sparsity pattern of the Slater matrix for a typical
configuration with $1024$ particles and $k=1$. As the basis functions have local support
(after truncating outside a cut-off radius) and the electrons, on average, are
distributed evenly, the sparsity pattern is similar to that of a finite difference matrix
for a regular 3D grid and a relatively wide stencil, where the width of the stencil is comparable
to the cut-off radius of the Gaussian orbitals (which in turn depends on $k$).
Second, we give typical spectra for matrices arising from systems
with $1024$ particles and $k= 1.5$, $1$, and $0.5$. We point out that
for the physics the ordering of rows and columns is irrelevant, since the square of the
determinant is invariant under such changes. However, the eigenvalues can change
significantly under reordering of rows and columns, which in turn can have a significant
influence on the convergence of iterative methods (see section~\ref{sec:AlgoImprove} for
the reordering algorithm used). In figures~\ref{figure:spec_k=15}--\ref{figure:spec_k=05},
for each value of $k$, we provide the spectrum of a
matrix before reordering, the spectrum of that same matrix after reordering, and the spectrum
of that matrix after reordering and with preconditioning.
The latter spectrum is the most relevant for the iterative solver.

Next, in Table~\ref{tab:Cond_k}, we provide typical condition numbers for various values
of the parameter $k$ (in the Gaussian orbitals) and three problem sizes, $n=686$, $1024$,
and $2000$. Note that, although
for $k=0.5$ the condition number appears to increase slightly by preconditioning,
the spectrum improves drastically. This also bears out in the iterative solver;
preconditioning reduces the number of iterations. From Table~\ref{tab:Cond_k}, we see that the
condition number of the unpreconditioned Slater matrix increases with decreasing $k$. This is
to be expected; see, e.g., \cite{BoyGill11}.
\begin{table}[h]\label{tab:Cond_k}
\begin{tabular}{|c|ccc|ccc|ccc|}
  \hline
  $k$ & \multicolumn{3}{|c|}{$1.5$} &
        \multicolumn{3}{|c|}{$1.0$} &
        \multicolumn{3}{|c|}{$0.5$} \\ \hline
  Size & 686 & 1024 & 2000 & 686 & 1024 & 2000 & 686 & 1024 & 2000 \\ \hline
  $\k(A)$ &
    73      & 1.4e2 & 1.3e3 &
    1.6e2   & 6.7e2 & 6.7e2 &
    2.4e3   & 1.1e4 & 8.3e3 \\
  $\k\left( A (LU)^{-1} \right)$ &
    1.2     & 1.4   & 4.7 &
    3.1     & 10    & 23  &
    8.0e3   & 1.5e5 & 1.1e5 \\
  \hline
\end{tabular}
\caption{Typical (spectral) condition numbers, $\k$, for Slater matrices and ILUTP preconditioned Slater matrices for $k = 1.5$, $1$, $0.5$ and problem sizes $n = 686$, $1024$, $2000$.}
\end{table}

Although the orbitals for realistic systems (physical materials) may differ
significantly from Gaussians, we expect many of the properties of the resulting
Slater matrices to be similar. If orbitals decay sufficiently fast, the matrix will
have the same banded sparsity pattern (after appropriate
reordering) and be diagonally dominant or nearly so. In that case, all or most eigenvalues
will be in the right half plane. However, a poor ordering of rows and columns will
lead to eigenvalues located around the origin, as is the case here.
If the decay is slow the matrix will become more
ill-conditioned. Moreover, if the decay is sufficiently slow,
there will be no ordering that yields diagonal
dominance and the spectrum cannot be guaranteed to be in the right-half plane. In that case,
we expect that there will be no ordering that, by itself, will lead to a nice
spectrum (for iterative solvers),
and preconditioning will be more important. Analyzing these properties and their
dependency on the properties of orbitals, decay rate, and lattice type will be
future work.

\subsection{Markov Chain Monte Carlo}
The Monte Carlo method is used to compute the high dimensional
integrals of (\ref{expectationEnergyVMC}) below. Direct
sampling would be very inefficient, because the wave function
assumes large values only in a small region of the $3n$
dimensional space. Therefore, a Markov Chain Monte Carlo
algorithm (MCMC) using a Metropolis update rule is used. For a
comprehensive discussion of the Metropolis algorithm
see~\cite{CepChes77,foulkes:review,thijssen:book}. The first
VMC for bosonic systems was reported in \cite{McMi65}.

% Add McMillan paper
% W.L. McMillan Phys. Rev. A 138, 442 (1965)  First VMC for bosonic systems.

The MCMC algorithm samples configurations as follows. At each
step, the current configuration $R$ (representing the
collective coordinates of all particles) is changed by moving
one particle a small (random) distance, generating a trial
configuration $R'$. The particles can be moved in order or by
random selection.
%In addition, it is possible to move multiple
%or even all particles in a single step, but in this paper we
%focus on moving the particles one by one.
The trial configuration is accepted with a probability that is
equal to the ratio of the probabilities (densities) of the two
configurations, assuming uniform sampling of the trial
coordinate. Hence we compute $\frac{|\Psi(R')|^2}{|\Psi(R)|^2}$
and compare with a random number drawn from the uniform
distribution on $(0,1)$; the new configuration is accepted if
the ratio is larger than the random number. Hence, this is
where the determinant ratios arise. The exponentials of Jastrow
factors must be computed as well, but since they are cheap for
sufficiently large $n$ (in fact, they are identical in form to
what is done in well-studied classical simulations
\cite{FreSmi02,AllTildes87}), we will
ignore them in this paper. If the trial configuration is
accepted, the new configuration becomes $R'$, otherwise the new
configuration is $R$ (again). Since we move a single particle
at each step, say particle $i$, in the trial configuration only
$r_i$ is changed to $r'_i$, and the Slater matrices
(\ref{eq:SlaterMatrix}) for the current configuration $R$ and
the trial configuration $R'$ differ only in row $i$. Therefore,
consecutive Slater matrices in our MCMC algorithm differ in one
row or are the same (when the trial configuration is rejected).
We refer to the attempted move of one particle as a step and to
the sequence of attempted moves of all particles as a
sweep.\footnote{Multiple or all particle moves can also be made
but require more sweeps (because the rejection rate is higher)
and are no more efficient per sweep.}

\subsection{Local Energy}
One important property of many body systems to calculate is the
expectation value of the energy  \cite{foulkes:review},
\begin{equation}\label{expectationEnergyVMC}
    E_V = \frac{\int \Psi_{\alpha}^{*}(R) H \Psi_\alpha(R) dR}
    {\int \Psi_{\alpha}^{*}(R) \Psi_\alpha(R) dR}
        = \frac{\int |\Psi_{\alpha}(R)|^2 E_{L,\a}(R) dR}{\int
            \Psi_{\alpha}^{*}(R) \Psi_\alpha(R) dR} ,
\end{equation}
where $E_{L,\a}(R) = (H \Y_{\a}(R))/\Y_{\a}(R)$ is referred to
as the local energy, and $H$ denotes the Hamiltonian of the
system; see, for example, \cite[section 4.5]{thijssen:book} and
\cite[p. 45]{Ha94}. Notice that the observable $E_{L,\a}(R)$
averaged over samples taken from the VMC sampling gives us the
expectation value of the energy. The algorithm assumes that
$\Psi_\alpha(R)$ and $\nabla \Psi_\alpha(R)$ are continuous in
regions of finite potential. The computation of $E_{L,\a}(R)$
requires the calculation of the Laplacian $\nabla^2_i
\Psi_\alpha(R)/\Psi_\alpha(R)$ and gradient $\nabla_i
\Psi_\alpha(R)/\Psi_\alpha(R)$ with respect to each particle
$i$.  This must be done once per sweep. These quantities are
computed by evaluating another determinant ratio. For a given
$i$, we replace row $i$ in the Slater matrix by its Laplacian
respectively its gradient, and then evaluate the ratio of the
determinant of this matrix with the determinant of the Slater
matrix. As this has to be done once per sweep, it also scales
as $O(n^3)$, and the methods described in this paper naturally
generalize to evaluating these quantities by iterative methods
with a reduced scaling.

As a point of note, the only computationally slow aspect of the
Diffusion Monte Carlo method (DMC) that differs from VMC
involves computing the gradient of the wave function for
particle $i$ at each step where particle $i$ is moved. Again,
the determinant ratio methods discussed in this paper are also
applicable to this situation.

% It should be pointed out that when evaluating our
%new method by comparing the total energies, we used the
%standard (and hence slow to calculate) method to evaluate the
%local energy so as to not confound the errors in using the
%approximation twice.
% It also assumes that the two integrals in (\ref{expectationEnergyVMC}) exist.
% The above steps demonstrate the application of the variational principle.

\subsection{Optimization}
The outer loop of the VMC method consists of updating the
vector of variational parameters, $\alpha$, after the MCMC
evaluation of the average of $E_{L,\alpha}$, so as to minimize
the total average energy. In principle, these variational
parameters could vary over all possible functions for all $n$
single particle orbitals.  In practice, the orbitals are often
optimized over a smaller subclass of possible functions.  For
example, in our model system, one might imagine optimizing $k$
or the location of the orbitals, $Z_j$ (i.e., off the b.c.c.
lattice). In more realistic scenarios, each orbital itself has
more structure and might be expanded in terms of basis
functions such as plane waves or Gaussian basis functions
(\ref{gaussian}).  Often, then, the expansion factors will be
optimized (in addition to the Jastrow factors).  Care must be
taken when using more complicated representations. If, for
example, each single particle orbital is composed of $n$ plane
waves and there are $n$ orbitals to be evaluated for $n$
particles, even constructing the Slater matrix would scale as
$O(n^3)$ per sweep. However, by tabulating the orbitals on a
grid, and doing a table lookup when needed, the cost of
constructing the matrix can be brought down to $O(n^2)$
operations per sweep, since there are $O(n^2)$ matrix elements,
or $O(n)$ if the matrix is sparse.

In the next section, we describe our algorithm, which reduces
the cost of evaluating determinant ratios
for physically realistic systems to about $O(n^2)$ per
sweep.

\section{Algorithmic Improvements}\label{sec:AlgoImprove}

As described in the previous section, the sequence of particle
updates, moving particle $i_k$ at step $k$ in the MCMC method,
leads to a sequence of matrix updates for the trial
configuration,
\begin{equation}\label{eq:matrixUpdate2}
    \tilde{A}_{k} = A_{k} + e_{i_{k}} u_{k}^T
                  = A_k (I + A_{k}^{-1} e_{i_k} u_k^T),
\end{equation}
where
\begin{equation}\label{matrixUpdateU}
    (u_{k})_j = \phi_j(r'_{i_{k}}) - \phi_j(r_{i_{k}}), \quad
    \mbox{ for } j = 1, \ldots, n,
\end{equation}
$A_{k}$ is the Slater matrix at the ${k}^{th}$ Monte Carlo
step, $e_{i_{k}}$ is the Cartesian basis vector with a $1$ at
position $i_{k}$, the $\phi_j$ are the single particle orbitals
used in (\ref{eq:WaveFunction})--(\ref{eq:SlaterMatrix}), and
$r_{i_{k}}$ and $r'_{i_{k}}$ are the old and the new position
of the particle $i_{k}$, respectively. We do not need to
compute $\phi_j(r_{i_{k}})$ since it equals
$(A_{k})_{i_{k},j}$. The acceptance probability of the trial
configuration depends on the squared absolute value of the
determinant ratio of the two matrices,
\begin{equation}\label{eq:detRatioEq}
    \frac{| \det \tilde{A}_{k} |^2}{| \det A_{k} |^2} =
      \left| 1 +  u_{k}^T A_{k}^{-1} e_{i_{k}}\right|^2 ,
\end{equation}
which can be computed by solving the linear system $A_k z_k =
e_{i_k}$ and taking the inner product $u_{k}^{T}z_k$. We
compare the value from (\ref{eq:detRatioEq}) with a random
number drawn from the uniform distribution on $(0,1)$. If the
trial configuration giving $\tilde{A}_k$ is accepted, $A_{k+1}
= \tilde{A}_k$, if the trial configuration is rejected,
$A_{k+1} = A_k$.

The use of maximally localized single particle orbitals leads
to a sparse Slater matrix in some cases (insulators).
%In this case, preconditioned iterative solvers can be very efficient.
%Preconditioning means that we solve $A_k M_k \hat{z}_k =
%e_{i_k}$, with preconditioner $M_k$ and $z_k = M_k \hat{z}_k$.
In this case, iterative solvers provide a promising approach to
compute these determinant ratios, as long as effective
preconditioners can be computed or updated cheaply. Variations
of incomplete decompositions (such as incomplete LU) are good
candidates for this problem, as they have proven effective for
a range of problems and require no special underlying structure
(in principle).
%However, an incomplete decomposition may not exist,
Unfortunately, the sequence of particle updates leads to
matrices that are far from diagonally dominant, often have
eigenvalues surrounding the origin, and have unstable
incomplete decompositions in the sense defined in
\cite{benzi:survey,SaadBk03}. However, the properties of
orbitals and localization suggest that with a proper ordering
of orbitals and particles the Slater matrix will be nearly
diagonally dominant. Our method resolves the preconditioning
problem by combining the following three improvements.

First, we have derived a geometric reordering of electrons and
orbitals, related to the ordering proposed in
\cite{edmunds:io}, that provides nearly diagonally dominant
Slater matrices. This ordering combined with an ILUTP
preconditioner \cite{Saad94,SaadBk03} leads to very effective
preconditioners; see section~\ref{sec:expmnts}. However,
reordering the matrix and computing an ILUTP preconditioner
every (accepted) VMC step would be too expensive. Therefore, as
the second improvement, we exploit the fact that $A_{k+1} = A_k
(I + A_{k}^{-1} e_{i_k} u_k^T)$ (\ref{eq:matrixUpdate2}) and
use a corresponding update to the right preconditioner, $M_k$,
such that $A_{k+1}M_{k+1} = A_k M_k$. This leads to cheap
intermediate updates to our preconditioners. Moreover, if $M_k$
has been computed such that $A_k M_k$ has a favorable spectrum
for rapid convergence \cite{VorsBk03,GreeBk97,SaadBk03}, then
subsequent preconditioned matrices, $A_{k+s}M_{k+s}$, have the
same favorable spectrum. However, each update increases the
cost of applying the preconditioner, and we periodically
compute a new ILUTP preconditioner. Third, we assess whether
potential instability of the incomplete LU decomposition
affects the iteration by applying the stability metrics from
\cite{benzi:survey,SaadBk03,chow:vector} in an efficient
manner. We only reorder the matrix if instability occurs, or if
our iterative solver does not converge to the required
tolerance in the maximum number of iterations, or if our iterative
solvers takes more than four times the average number of
iterations. This monitoring
and reordering is necessary as the matrix becomes far from
diagonally dominant and the incomplete LU decompositions slowly
deteriorate due to the continual updates to the Slater matrix.
This approach has proved very effective and limits the number
of reorderings to a few per sweep.
% Hence reordering does not affect the scaling of our method.
In spite of the matrix reordering, the explicit reordering of electrons and
orbitals, based on the stability metric discussed
below or on slow convergence, pivoting in the incomplete LU
factorization is necessary. If we do not pivot the factorization
occasionally breaks down. Such a breakdown could
be avoided by doing the explicit reordering of electrons and orbitals
every step, but this would be too expensive. Moreover, not pivoting
leads to denser $L$ and $U$ factors and slower
convergence in the iterative linear solver; both effects increase
the total amount of work.

\subsection{Reordering for Near Diagonal Dominance and Efficient Reordering Criteria}\label{reorder}
%The continual changes in the matrix lead to matrices with poor
%ILUTP factorizations, and we have to reorder from time to time.
%However, since reordering is expensive we must limit the number
%of reorderings to the minimum possible.
%The quality of the preconditioner, yielding fast convergence,

First, we discuss an efficient way to judge the quality of the
preconditioner. In the second part of this section, we discuss
a reordering that improves the quality of the preconditioner.

%% Part A - Accuracy and Stability
The quality of an ILU preconditioner for the matrix $A$, $LU
\approx A$, can be assessed by its {\em accuracy} and {\em
stability} \cite{benzi:survey}. The accuracy of the ILU
preconditioner, defined as
\begin{equation}\label{accuracyEq}
    N_1 = ||A - LU||_F ,
\end{equation}
measures how close the product $LU$ is to the matrix $A$. The
stability of the preconditioner, for right preconditioning
defined as
\begin{equation}\label{stabilityEq}
    N_2 = ||I - A(LU)^{-1}||_F ,
\end{equation}
measures how close the preconditioned matrix is to the
identity. For left preconditioning, $N_2 = ||I -
(LU)^{-1}A||_F$. Although for some classes of matrices $N_1$ is
a good measure of preconditioning quality, for general matrices
$N_2$ is a more useful indicator
\cite{benzi:survey,benzi:orderings}. We will see that this is
also the case here. In general, instability arises from a lack
of diagonal dominance and manifests itself in very small pivots
and/or unstable triangular solves.

In practice, computing $N_2$ is much too expensive, and we need
to consider a more economic indicator. An alternative approach,
suggested in~\cite{chow:vector}, is to compute
$||(LU)^{-1}e||_\infty$, where $e$ is the vector of all $1$'s.
However, this still requires solving an additional linear
system. Instead, we propose to use an {\em effective} or {\em
local} stability measure,
\begin{equation}\label{stabilityVecEq}
    N = \max_i \|v_i - A(LU)^{-1}v_i\|_2,
\end{equation}
where the $v_i$ are the Arnoldi vectors generated in the GMRES
algorithm \cite{SaadSchu86} during a linear solve. $N$ measures the instability
of the preconditioned matrix over the Krylov space from which
we compute a solution. If $N$ is small, there is no unit vector
$z \in K_m(A(LU)^{-1},r_0)$ for which $\|(I - A(LU)^{-1})z\|_2$
is large (where $m$ is the number of GMRES iterations). Indeed,
for unit $z \in K_m(A(LU)^{-1},r_0)$, $\|(I - A(LU)^{-1})z\|_2
\leq mN$. $N$ can be small or modest when $N_2$ is large;
however, this indicates that the instability does not play a
role for vectors in the subspace over which a solution was
computed; hence, the name effective or local stability. Note
that $N$ can be computed during the GMRES iteration at
negligible cost, in contrast to the expensive computation of
$N_2$. If $N$ is large, the preconditioned matrix is
ill-conditioned over the Krylov space used to find a solution,
and we reorder the matrix as described below. Large $N$
indicates that the solution might be inaccurate and that the
preconditioner is deteriorating. This typically would lead
to poor convergence either in the present solve or later, and 
so reordering and updating the preconditioner is better.

To check whether $N_1$ and $N_2$ are good indicators of
preconditioner quality for our problem, we run the MCMC
algorithm for $100$ sweeps for a test problem with $2000$
particles, and we check $N_1$ and $N_2$ each time GMRES does
not converge in $15$ iterations (which is relatively slow; see
section~\ref{sec:expmnts}). It would be useful to check the
reverse as well, but computing $N_1$ and $N_2$ for every MC
step ($200,000$ steps) would be too expensive. If GMRES does
not converge in $15$ iterations, we reorder the matrix as
described below, compute a new ILUTP preconditioner, and solve
the linear system again from scratch.
This procedure always led to convergence
within $15$ iterations after the reordering. Although the
experiments in section~\ref{sec:expmnts} are computed using a
C/C++ code, for experiments in this subsection we use a Matlab
based VMC code, developed for easy experimentation, that uses
the {\tt GMRES} and {\tt luinc} routines of Matlab. The code
uses left preconditioning, {\tt luinc} with a drop tolerance of
$\tau = 0.01$, and it allows pivoting in the entire pivot
column (default). Furthermore, we use Gaussian orbitals
(\ref{gaussian}) with $k = 1$.
\begin{figure}[bt]
    \centering
    \includegraphics[scale=0.50]{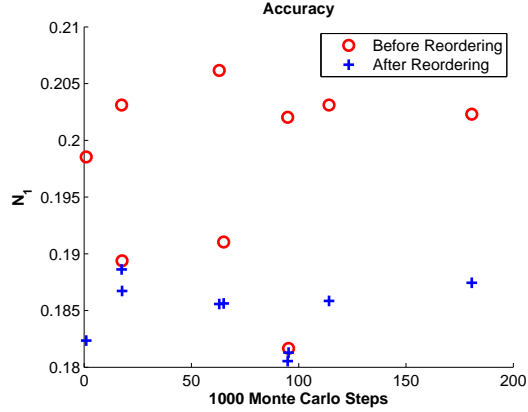}
    \caption{Accuracy before and after reordering.}
    \label{figure:accuracy}
\end{figure}
\begin{figure}[tb]
    \centering
    \includegraphics[scale=0.50]{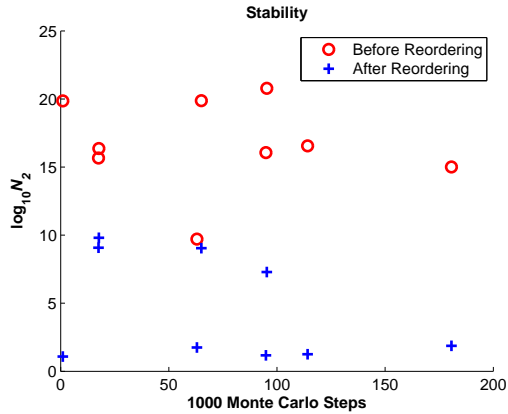}
    \caption{Stability before and after reordering.}
    \label{figure:stability}
\end{figure}
In Figures~\ref{figure:accuracy} and \ref{figure:stability} we
plot, respectively, $N_1$ and $N_2$ at those Monte Carlo steps
where the GMRES algorithm does not converge in $15$ iterations,
implying a deterioration of the preconditioner. We also plot
$N_1$ and $N_2$ after reordering and recomputing the
preconditioner. Figure~\ref{figure:accuracy} shows that the
accuracy is always quite good, and hence accuracy is not a good
indicator for reordering the matrix to improve preconditioner
quality. This is in line with observations from
\cite{benzi:survey}. Reordering does improve the accuracy
further. Figure~\ref{figure:stability} shows that poor
convergence goes together with very large values of $N_2$, and
so $N_2$ is a better indicator for reordering the matrix to
improve the preconditioner. Note that reordering improves the
stability significantly, and usually reduces it to modest
values, but with some exceptions. As $N_2$ seems a good
indicator, but too expensive, we next consider the effective
stability $N$.
\begin{figure}[bt]
    \centering
    \includegraphics[scale=0.50]{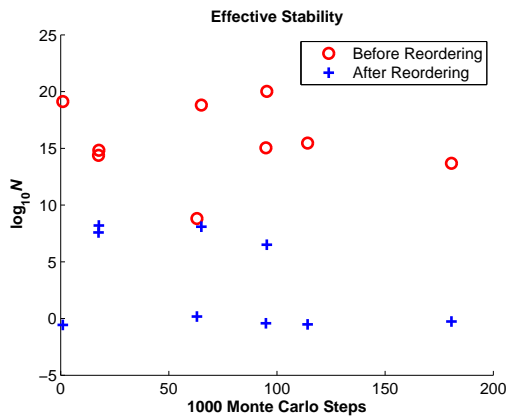}
    \caption{Effective stability before and after reordering.}
    \label{figure:effective}
\end{figure}
Figure~\ref{figure:effective} shows large values of $N$
corresponding to steps where the GMRES algorithm does not
converge. The value of $N$ is modest at other Monte Carlo steps
(note that it is easy to compute $N$ at every step). This
demonstrates that $N$ is an equally good indicator of the
effectiveness of the preconditioner as $N_2$, and hence we will
use $N$ as an indicator for matrix reordering. We remark that
more timely reordering (before $N$ gets so large that the
matrix is numerically singular) also leads to smaller values
for $N$ after reordering the matrix and recomputing the
preconditioner. Next, we discuss reordering the matrix when
the quality of the preconditioner deteriorates.

%% Part B - Reordering to improve effectiveness of the preconditioner
Instability in the incomplete factorization, as discussed above
(small pivots and/or ill-conditioned triangular solves), is
generally associated with a lack of diagonal dominance
\cite{chow:vector,benzi:survey}. Instability can generally be
removed or significantly reduced by preprocessing the
coefficient matrix. Preprocessing strategies involve
permutations and scalings for improving the conditioning,
near-diagonal dominance, and the structure of the coefficient
matrix; modifications such as perturbing diagonal elements are
also possible \cite{chow:vector,benzi:survey}. Improving
near-diagonal dominance of the matrix through a judicious
ordering of unknowns and equations has been shown quite
effective in addressing instability, leading to a better
preconditioner. In~\cite{benzi:orderings}, the authors show
that a simple reordering of the matrix can improve the quality
of incomplete factorization preconditioners. More sophisticated
reorderings that improve the near-diagonal dominance of the
matrix are discussed in~\cite{duff:diag, duff:diag2}. These
papers also show that reordering can have a dramatic effect on
the convergence of ILU preconditioned iterative methods.
Reorderings that exploit the physics of the underlying problem
have also proved quite effective \cite{chernesky:physics}.

We remark that the physics underlying our problem and the
optimization of orbitals suggest that a proper ordering should
lead to a nearly diagonally-dominant matrix. We also observe
that, as all orbitals are scaled equally, we do not expect
scaling to provide much improvement.

%DMC
Since the orbitals used in our study are monotonically
decreasing with distance, we propose a reordering that is
simple, improves the near-diagonal dominance of the Slater
matrix, and incorporates the physics of our problem. This
reordering strategy performs a geometric reordering of
particles and orbitals, and is similar to the reordering of
inputs and outputs for a reliable control scheme
in~\cite{edmunds:io}. Our algorithm consists of the following
steps, ignoring sparsity for simplicity.
\begin{tabbing}
  xx \= xx \= xx \= xx \= xx \=  \kill
  % \> for next tab, \\ for new line...
  1. \> Label the particles ($P_i$) and orbitals ($O_j$) from $1$
        to $n$, \\
     \> giving the following Slater matrix $A$: \\
     \>  \begin{minipage}[c]{3in}
           \begin{align*}\label{firstMat}
           \bordermatrix{%
                 &  O_1 &  O_2 & & O_n \cr
             P_1 & \phi_1(r_1) & \phi_2(r_1) & \cdots & \phi_n(r_1)  \cr
             P_2 & \phi_1(r_2) & \phi_2(r_2) & \cdots & \phi_n(r_2)  \cr
                 & \vdots & \vdots &  & \vdots \cr
             P_n & \phi_1(r_n) & \phi_2(r_n) & \cdots & \phi_n(r_n)  \cr
           },
           \end{align*}
     \end{minipage} \\
     \> \\
  2. \> for $i=1$, \ldots, $n-1$ do  \\
     \> \> Find the closest orbital $O_j$ to $P_i$ for $j \in \{i, i+1, \ldots, n\}$\\
     \> \> if $j \neq i$ then  \\
     \> \> \> renumber $O_j$ as $O_i$ and $O_i$ as $O_j$ (swap columns $j$ and $i$) \\
     \> \> else \\
     \> \> \> find the particle $P_k$ closest to orbital $O_i$
              for $k \in \{i, i+1, \ldots, n\}$ \\
     \> \> \> if $k \neq i$ then \\
     \> \> \> \> renumber $P_k$ as $P_i$ and $P_i$ as $P_k$
                 (swap rows $i$ and $k$) \\
     \> \> \> end if \\
     \> \> end if \\
     \> end for \\
\end{tabbing}

As mentioned above, reordering the matrix and recomputing the
ILUTP preconditioner always leads to convergence within $15$
GMRES iterations. So, the algorithm is quite effective. We can
see from Figure~\ref{figure:effective} that $N$ is always
significantly reduced by reordering the matrix and recomputing
the preconditioner. Moreover, more timely reordering of the
matrix, before $N$ gets so large that the matrix is numerically
singular, also leads to smaller values for $N$ after
reordering.
%Furthermore, in
%\cite{chow:vector} it is pointed out that problems with
%preconditioners mainly arise when the some of the (their)
%metrics are very large.

The computational cost of a straightforward implementation of
this reordering is $O(n^2)$. Since for our current experiments
(and problem sizes) the runtime of reordering is negligible and
reordering is needed only two or three times per sweep, we have
not (yet) focused on an efficient implementation, especially
since this likely requires substantial work in the software for
matrix storage and manipulation. However, we will do this in
the future. We remark that for sparse matrices where the number
of nonzeros per row and per column is roughly constant,
independent of $n$, an $O(n)$ implementation is possible.
%However, this is due to the matrix storage scheme,
%compressed sparse row (CSR), which makes searching the matrix
%row-wise very efficient but expensive column-wise. The reason
%for the storage scheme is the efficient matrix-vector product
%and computation of the preconditioner. A better storage scheme
%would reduce the cost of the reordering (for a sparse matrix)
%to $O(n)$.
%As the actual cost of the reordering is modest (and
%typically $O(n)$), we leave this for future work.
Moreover, this global reordering algorithm ignores the local
nature of the particle updates and resulting changes to the
matrix. In addition, maintaining further (multilevel) geometric
information related to relative positions of particles and
orbitals should make reordering more efficient. Hence, longer
term we expect to replace this algorithm by one that makes
incremental updates to the local ordering. If we start with a
good ordering and the matrix is sparse (in the sense that the
decay of the orbitals does not depend on the problem size), we
expect that such local updates can be done at constant or near
constant cost.

\subsection{Cheap Intermediate Updates to the Preconditioner}\label{updatePrecond}

Computing a new ILUTP preconditioner for every accepted
particle update would be very expensive. However, we can
exploit the structure of the matrix update to compute a cheap
update to the preconditioner that maintains good convergence
properties.

Assume that at some step $k$ we have computed an incomplete LU
preconditioner (ILU) with threshold and pivoting from the
matrix $A_k$. We have
\begin{equation}\label{ilu}
    A_k Q_k \simeq L_k U_k ,
\end{equation}
where $Q_k$ is a column permutation matrix, and $L_k$ and $U_k$
are the incomplete lower and upper triangular factors,
respectively. We consider right preconditioning, and so,
instead of $A_k z_k = e_{i_k}$, we solve the right
preconditioned linear system,
\begin{equation}\label{precondSystem}
    A_k M_k \tilde{z}_k = e_{i_k}
    \quad\mbox{with}\quad z_k = M_k \tilde{z}_k,
\end{equation}
where $M_k = Q_k \left(L_k U_k\right)^{-1}$ is the
preconditioner. If the trial move of particle $i_k$ is
accepted, we consider for the next step the updated matrix
$A_{k+1} = A_k (I + (A_{k}^{-1} e_{i_k}) u_k^T)$; see
(\ref{eq:matrixUpdate2}). Now, let $M_k$ be such that $A_k M_k$
has a favorable spectrum for rapid convergence
\cite{VorsBk03,GreeBk97,SaadBk03}. Then defining the updated
preconditioner $M_{k+1}$ such that $A_{k+1}M_{k+1} = A_k M_k$
gives a new preconditioned matrix with the same favorable
spectrum. Hence, we define the new preconditioner as
\begin{equation}\label{finalMk0}
    M_{k+1} = (I + (A_{k}^{-1} e_{i_k}) u_k^T)^{-1}M_k =
    \left(I - \frac{(A_k^{-1}e_{i_{k}}) u_{k}^T}{1+u_{k}^T A_k^{-1}e_{i_{k}}}
    \right)M_k ,
\end{equation}
without explicitly computing $M_{k+1}$. Since
$A_k^{-1}e_{i_{k}} = z_k$ and $1+u_k^T A_k^{-1}e_{i_{k}}$ have
already been computed to find the determinant ratio
(\ref{eq:detRatioEq}), we get $M_{k+1}$ for free, and the cost
of applying $M_{k+1}$ is that of applying $M_k$ plus the cost
of a dot product and vector update. Let $\hat{z}_k = (1+u_k^T
A_k^{-1}e_{i_{k}})^{-1} z_k$. Then $M_{k+1}$ is defined as
\eqs%
\label{eq:mkp1}
  M_{k+1} & = & (I - \hat{z}_k u_k^T)Q_k(L_k U_k)^{-1} ,
\eqe%
where the inverse of $L_k U_k$ is implemented, of course, by a
forward solve for $L_k$ and a backward solve for $U_k$.

Since $z_k$ is approximated by an iterative process, and there
is no need to solve very accurately, we have $A_{k+1}M_{k+1}
\approx A_k M_k$, rather than exact equality. However, we have
the following result. Let $\z_k = A_k^{-1}e_{i_k}$ and let $r_k
= e_{i_k} - A_k z_k = A_k(\z_k - z_k)$. Then we have
\eqsn%
  A_{k+1}M_{k+1} & = & A_k(I + \z_ku_k^T)(I -
  \frac{z_ku_k^T}{1+u_k^Tz_k})M_k \\
  & = & A_k \left(I + \z_ku_k^T - \frac{\z_ku_k^T}{1+u_k^Tz_k} +
        \frac{(\z_k - z_k)u_k^T}{1+u_k^Tz_k} -
        \z_ku_k^T \frac{u_k^Tz_k}{1+u_k^Tz_k} \right)M_k \\
  & = & A_k \left(I + \frac{(\z_k -
        z_k)u_k^T}{1+u_k^Tz_k}\right) M_k  \\
  & = & A_kM_k + \frac{r_k u_k^T}{1+u_k^Tz_k}M_k .
\eqen%
So, the relative change in the preconditioned matrix is small
unless $|1+u_k^Tz_k|$ is very small or $\|r_k\|\|M_k\|$ is
large relative to $\|A_kM_k\|$. However, $|1+u_k^Tz_k|$ governs
the acceptance probability of the particle move. So, a very
small value would occur in the preconditioner update with very
small probability; we do not need to update the preconditioner
if a trial move is rejected. This also guarantees that an
accepted particle move will never result in a singular matrix,
as the move will be accepted with probability $0$. If $\|M_k\|$
is large while $\|A_kM_k\|$ is small (say $O(1)$), then $A$
must have small singular values. In that case, an accurate
$z_k$ requires a sufficiently small residual. Hence, unless
$\|A_kM_k\|$ is large (which monitoring $N$ guards against), a
proper choice of stopping criterion for $\|r_k\|$ should keep
the relative change in the preconditioned matrix small.

Obviously, we can repeat the update of the preconditioner for
multiple updates to the matrix. Defining $z_s =
A_s^{-1}e_{i_s}$ (approximately from the iterative solve) and
$\hat{z}_s = (1+ u_s^Tz_s)^{-1} z_s$, where $u_s$ is given by
(\ref{matrixUpdateU}) and $s = k, \ldots, k+m$, we have
\eqs%
  M_{k+m} & = & \left(I - \hat{z}_{k+m-1}u_{k+m-1}^T \right) \cdots
                \left(I - \hat{z}_{k+1}u_{k+1}^T \right)
                \left(I - \hat{z}_ku_{k}^T \right) M_k. \label{finalMkm}
\eqe%
In this case, the cost of applying the preconditioner slowly
increases, and we should compute a new ILUTP preconditioner
when the additional cost of multiplying by the multiplicative
updates exceeds the (expected) cost of computing a new
preconditioner. Of course, we also must compute a new
preconditioner if we reorder (large $N$).

This technique for updating the preconditioner is similar to
the idea of updating the (exact) inverse of the Jacobian matrix
for Broyden's method \cite{kelley:book}, which is applied to
the exact inverse of the Slater matrix in \cite{nukula:slater}.
See also \cite{BerBru06} which uses a similar approach to
updating a preconditioner, however, for a general nonlinear
iteration.
% \emph{QUESTION:  WHAT DOES THIS SENTENCE MEANS?}
So, the notion to keep the preconditioned matrix fixed,
$A_{k+1}M_{k+1} = A_kM_k$, appears to be new.

\section{Numerical Experiments}\label{sec:expmnts}

%These choices
%are selected to be physically relevant so that
%the results give confidence that realistic problems can be
%addressed by our method; yet the model problem is sufficiently
%simple that we can easily understand the behavior of the
%algorithm and test relevant statistics.

In this section, we numerically test our new algorithm.
Apart from testing the performance of our algorithm, we must
also test its reliability and accuracy. Since we compute the
determinant ratios in the MCMC algorithm by iterative methods,
we replace the acceptance/rejection test by an approximate
test. In addition, the updating of preconditioners and their
dependence on occasional reordering may lead to slight
inconsistencies in the computed acceptance/rejection
probabilities. This may affect the property of detailed balance
for equilibrated configurations. Although we assume that using
sufficiently small tolerances would make such potential
problems negligible,
%and small unbiased errors do not affect
%the computational results,
we test and compare an observable (kinetic energy) computed in
the new algorithm and in the standard algorithm. We remark that
the standard method, with an updated inverse, also has the
potential of accumulation of errors, which may affect its
accuracy. However, the standard algorithm has performed
satisfactorily, and a successful comparison of the results of
our algorithm to those of the standard algorithm should
engender confidence.

As described in section~\ref{sec:vmc}, we use Gaussian
functions for the single-particle orbitals with $k=1$, and we
ignore the Jastrow factor. The lattice (giving the approximate
locations of the nuclei in a solid) is selected as a Body
Centered Cubic (b.c.c.) lattice, a lattice formed by cubes with
nuclei/orbitals on all vertices and one in the middle. To test
the scaling of our method, we fix the density of the electrons
at $(3/4\pi)=0.2387$ ptcl/unit and increase the number of
electrons $n$; this corresponds to increasing the number of
cubes in the lattice. This causes the number of orbitals to
increase linearly with the electron number. As the orbitals are
located on a b.c.c. lattice, we choose values of $n=2K^3$ where
$K$ is an integer representing the number of orbitals along one
side of the computational domain, and so the b.c.c. lattice is
commensurate with the periodic box. Note that this leaves the
spacing of the lattice (the distance between orbitals) the same
for all $n$. The length of the side of a cube is $2.031$, and
the nearest neighbor distance is $1.759$. The calculation of
properties on larger and larger lattices is a typical procedure
in QMC simulations to estimate the errors induced by not
simulating an infinite number of particles. It is important to
recognize that, although we are simulating an insulator, the
electrons are not confined to the neighborhood of a single
orbital and move around the entire box (hence the need for
occasional reordering of the matrix). Since we use Gaussian
orbitals, the Slater matrix has no coefficients that are
(exactly) zero. However, most of the coefficients are very
small and negligible in computations. Therefore, to make the
matrix sparse, we drop entries less in absolute value than
$10^{-5}$ times the maximum absolute value of any coefficient
in the matrix.
% , which for practical purposes was taken to be $1$ (maximum possible).
% ISN"T THIS EQUAL TO 1?  WAS THE CUTOFF REALLY DEPENDENT ON THE OTHER MATRIX ELEMENTS?
The number of non-zeros per
row of the matrix varies between $40$ and $50$.

We use QMCPACK~\cite{kim:qmc} for our simulations and, in its
standard version, for comparison. In order to efficiently
implement our iterative algorithms, we rewrote a significant
part of QMCPACK to handle sparse matrices \footnote{Note that
the standard algorithm does not (and cannot) exploit sparsity in the matrix,
except in computing and updating the Slater matrix itself.}. It
might be advantageous to work with sparse vectors as well
(sometimes referred to as sparse-sparse iterations) as the
right hand sides in (\ref{precondSystem}) are Cartesian basis
vectors. On the other hand, the preconditioner may quickly
destroy sparsity of the iteration vectors. This remains
something to test in the future. We integrated the new
components of our VMC algorithm in QMCPACK \cite{kim:qmc}
(written in C/C++); this includes the GMRES algorithm
\cite{SaadSchu86}, the ILUTP preconditioner \cite{Saad94}, our
algorithms to update the preconditioner by rank-one updates to
the identity (section \ref{updatePrecond}), our reordering
algorithm (section \ref{reorder}), and our test for instability
of the preconditioner (section \ref{reorder}).
% Our new code integrates the
% various components VMC, GMRES, the ILUTP preconditioner (based
%on the Sparskit implementation \cite{Saad}), updates the
%preconditioner (Section \ref{updatePrecond}), and performs
%reordering (Section \ref{reorder}) .

To simulate the system and gather statistics on the quality of
the results for the new method and on its performance, we carry
out $120$ sweeps. We discount the data corresponding to the
first $20$ sweeps, which are used to ensure that the system is
in equilibrium.
% before results are for the system in equilibrium.

\begin{figure}[bth]
    \centering
    \includegraphics[scale=0.40]{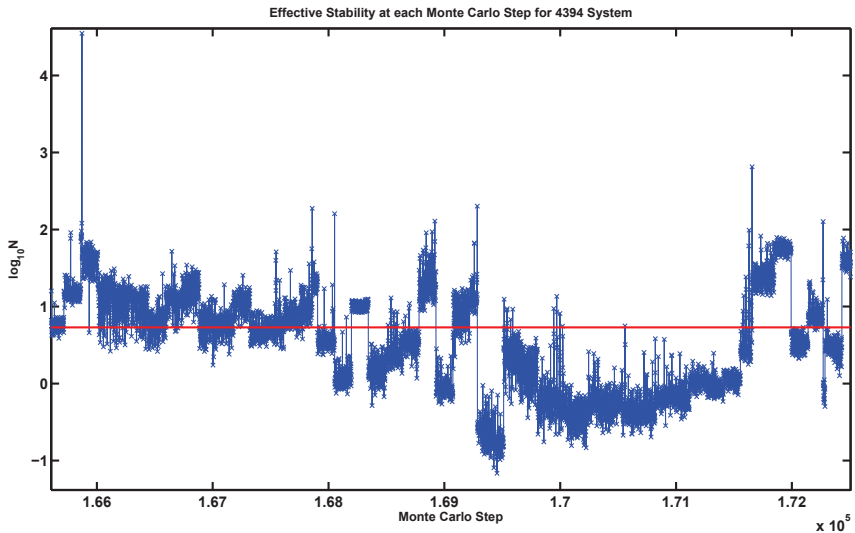}
    \caption{Effective stability, $N$, at each Monte Carlo step
    for the $4394$ system for a representative window of roughly $6000$ steps.
    The horizontal (red) line indicates the ($\log_{10}$ of) the average $N$ (the average $N$ is around $5.36$).}
    \label{figure:NvalsLog}
\end{figure}

\begin{figure}[tbh]
  \centering
  \includegraphics[scale=0.45]{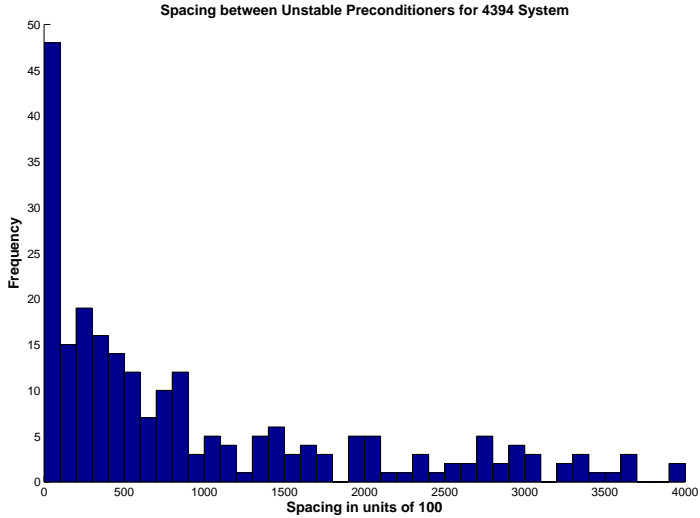}
  \caption{Histogram for the number of steps between reorderings (large $N$ or slow convergence)
  for the $4394$ system. The size of each bin is $100$.}
  \label{figure:SpacingHist}
\end{figure}

In the GMRES algorithm, we set the maximum number of iterations
to $40$, and the relative convergence tolerance to $10^{-6}$.
We monitor $N$, the effective stability (see section
\ref{reorder}), to decide when to reorder the matrix; we
reorder when $N > 100$. This is relatively low, but from
experience this leads to faster (average) convergence. Note
that, in our experiments, the number of reorderings is never
more than three per sweep (see below). Finally, since $N$ does
not always predict slow convergence, even though it is quite
reliable, we also reorder when the number of GMRES iterations
reaches four times the average number of iterations or when the
method does not converge in the maximum number of iterations.
However, the occurrence of slow convergence while $N$ is small
accounts only for a very small fraction of the total number of
reorderings, a few percent at most.
If $N > 100$ or GMRES does not converge fast enough,
we solve the same system again (from scratch) after
reordering and computing a new preconditioner (GMRES is not
restarted); this always resolves the convergence problem in our
tests. In figures~\ref{figure:NvalsLog} and \ref{figure:SpacingHist},
we provide some information on $N$ for the system with $4394$ particles.
In figure~\ref{figure:NvalsLog}, we give $\log_{10}(N)$ for each Monte Carlo step in
a representative window of roughly $6000$ MC steps (the number of reorderings in this 
window is slightly higher than average though);
in figure~\ref{figure:SpacingHist}, we provide a histogram for the number of
Monte Carlo steps between reorderings (each bin is of size $100$). We note that
sometimes reorderings follow each other with relatively short intervals. For
the $4394$ system, we have on average $2.73$ reorderings per sweep (see table~\ref{table:cost});
so, the number of steps between reorderings is, on average, about $1600$ steps.
However, we have a fair number of much shorter intervals. We conjecture that such shorter
intervals occur when multiple particles are not relatively close to any orbital (nucleus),
for example, when multiple pairs of electrons `swap' nuclei. As the matrix changes only by one row per
(successful) step and only by a modest amount, it is likely that matrices for which a good preconditioner
is difficult to compute are clustered in the MCMC sequence. Alternatively, this phenomenon might suggest that
we need a better reordering algorithm.
For preconditioning, we use the ILUTP preconditioner
from SPARSKIT~\cite{SaadBk03}. This preconditioner uses
pivoting and a relative threshold to drop (off-diagonal)
coefficients based on magnitude to obtain sparse lower
triangular and upper triangular factors, $L$ and $U$, such that
$AQ \approx LU$. We set the relative drop tolerance to $0.01$
and we allow pivoting in the entire row. We set the
permutation tolerance that governs the pivoting to $0.05$.
Finally, in ILUTP one can set the maximum number of additional
nonzeros (fill-in) per row allowed in both the $L$ and $U$
factor. We set this to half the average number of nonzeros per
row of the matrix, resulting in at most twice as many nonzeros
in the $L$ and $U$ factor together as in the matrix, $A$. The
average number of nonzeros in the preconditioner remains well
below this maximum; see Table~\ref{table:cost}.

The experimental results are given in four tables. In the first
two tables, we compare the results of our method to those of
the standard method to demonstrate the reliability and accuracy
of the new method. In the other two tables, we assess the
scaling of our method as the number of particles increases.

First, we assess how close the determinant ratio computed by
the new method is to that computed by the standard method. This
determines the probability of a {\em wrong} decision in the
acceptance/rejection test.  Let $q$ be the exact determinant
ratio squared and $q_a$ be the approximate determinant ratio
squared computed by an iterative method; then the probability
of a wrong decision at one step in the Metropolis algorithm is
$f = |min(q,1)-min(q_a,1)|$. The average value of $f$ over a
random walk (the entire sequence of MCMC steps), $\bar{f}$,
gives the expected number of errors in the acceptance/rejection
test (Exp. Errors in
Table~\ref{table:goodness}). %DMC
We call an approximation
\begin{itemize}
    \item extremely good if $f < 0.0001$,
    \item very good if $f < 0.001$, and
    \item good if $f < 0.01$.
\end{itemize}
The results of this test for eight successive problem sizes
($2\cdot 7^3$, $2\cdot 8^3$, \ldots, $2\cdot 14^3$) are given
in Table~\ref{table:goodness}. The high percentage of extremely
good approximations in Table \ref{table:goodness} shows that
approximating determinant ratios does not interfere with the
accuracy and reliability of the simulation. In fact, the new
algorithm makes a different accept/reject decision from that
made by the standard algorithm only once every $10^5$ steps.
Since the autocorrelation time is smaller than this by over an
order of magnitude, the system should quickly forget about this
`incorrect' step. Note that given the tolerance in GMRES the
step is not unlikely even in the standard algorithm. We also
report the acceptance ratio, which is defined as the ratio of
the number of accepts to the total number of Monte Carlo steps.
% is used to check the correctness of the Monte Carlo simulation.
The desired range of the acceptance ratio is between $0.2$ and
$0.8$, and this is satisfied in our simulation. Higher or lower
acceptance ratios are likely to create problems, as this
typically indicates that successive MC steps remain correlated
for a long time \cite{CalSom07}.

\begin{table}[th]
\small
    \begin{center}
        \begin{tabular}{|l|l|l|l|l|l|l|l|l|}
            \hline
            Size & 686 & 1024 & 1458 & 2000 & 2662 & 3456 & 4394 & 5488\\
            \hline
            Exp. Errors & 4.45e-6 & 4.22e-6 & 5.03e-6 & 4.41e-6 & 4.63e-6 & 4.50e-6 & 3.96e-6 & 4.07e-6 \\
            Extr. Good   & 99.49   &  99.53  &  99.57  &  99.49  &  99.50  & 99.47   & 99.56   & 99.56   \\
            Very Good   & 99.99   &  99.98  & 99.99   & 99.99   & 99.99   & 99.99   & 99.99   & 99.99   \\
            Good        & 100     & 100     & 100     & 100     & 100     & 100     & 100  & 100  \\
            Acc. Ratio  & 0.5879  & 0.5880  & 0.5887  & 0.5880  & 0.5881  & 0.5898  & 0.5878  & 0.5883  \\
            \hline
        \end{tabular}
    \end{center}
    \caption{Accuracy of the acceptance/rejection test, giving
    the average expected number of errors in the acceptance test (Exp. Errors), the percentage of
    extremely good approximations (Extr. Good), the percentage of very good
    approximations (Very Good), the percentage of good approximations (Good), and the acceptance ratio
    of trial moves (Acc. Ratio).}
    \label{table:goodness}
\end{table}

To further check the accuracy of the results of the simulation,
we compute the kinetic energy of the system (an important
observable). The kinetic energy of the system is defined as
\begin{equation}\label{energyEq}
    E = \frac{\hbar^2}{2\rho n} \sum_{ij} \left(6k - 4k^2||r_j - Z_i||^2\right)
    A_{ij}\left({A}^{-1}\right)_{ji},
\end{equation}
where $\hbar$ is the reduced Planck's constant ($h/2\pi$),
$\rho$ is the electron mass, $n$ is the system size, $k=1$ is
defined as in (\ref{gaussian}), $r_j$ is the position of
particle $j$, $Z_i$ is the position of orbital $i$, and $A$ is
the Slater matrix as given in (\ref{eq:SlaterMatrix}). (We use
units where $\hbar=\rho=1$). It should be noted that we use the
exact kinetic energy in this test even though, in practice, our
algorithm would also use iterative solvers to efficiently
evaluate this observable (which involves the Laplacian of the
wave function). This is done so as not to confound two sources
of errors (one being a few rare `incorrect' steps on the Markov
chain and the other being errors in computing the kinetic
energy). We compute the kinetic energy in two separate
experiments. However, in order to emphasize how close the
results of the new algorithm are to those of the standard
algorithm, {\em we start both experiments with the same
starting configuration and using the same initial seed for the
random number generator}. Since the expectation of an incorrect
acceptance/rejection is extremely small, both experiments
follow the same chain for an extended period of time, and
therefore the difference in kinetic energy between the
experiments is much smaller than the statistical variation that
would be expected if the chains were independent instead of
correlated. We compute the kinetic energy by sampling
(\ref{energyEq}) at the end of each sweep (doing $120$ sweeps
and discarding the first $20$). The first experiment computes
the determinant ratio for acceptance/rejection tests using the
standard QMCPACK algorithm, while the second experiment uses
the new (sparse) algorithm. The average energy over the whole
simulation is listed in Table \ref{table:energy}. It is evident
that the energies from the two algorithms are close for all
system sizes. We also compute the standard deviation of the
kinetic energy, $\s$, using DataSpork~\cite{ceperley:tool},
taking into account the autocorrelation\footnote{In Markov
Chain Monte Carlo, successive states tend to be correlated. The
autocorrelation measures how many steps of the algorithm must
be taken for states to be uncorrelated \cite{CalSom07}.}.

\begin{table}[h]
\small
    \begin{center}
        \begin{tabular}{|l|l|l|l|l|l|l|l|l|}
            \hline
            Size & 686 & 1024 & 1458 & 2000 & 2662 & 3456 & 4394 & 5488\\
            \hline
            \multicolumn{9}{|c|}{Standard QMCPACK Algorithm} \\
            \hline
            Energy              & 2.0984 & 2.1074 & 2.1107 & 2.1024 & 2.0964 & 2.0948 & 2.1035 & 2.1016 \\
            (hartree)           &&&&&&&&\\
            $\s$              &  0.0075 & 0.0077 & 0.0040 & 0.0045 & 0.0024 & 0.0028 & 0.0034 & 0.0035 \\
            \hline
            \multicolumn{9}{|c|}{Sparse Algorithm} \\
            \hline
            Energy              & 2.0984 & 2.1074 & 2.1107 & 2.1040 & 2.1010 & 2.0999 & 2.1049 & 2.1010 \\
            (hartree)           &&&&&&&&\\
            $\s$              & 0.0075 & 0.0077 & 0.0040 & 0.0034 & 0.0023 & 0.0045 & 0.0033 & 0.0034 \\
            \hline
        \end{tabular}
    \end{center}
    \caption{Kinetic Energy and Standard Deviation.}
    \label{table:energy}
\end{table}

Next, we analyze the performance of our algorithm and compare
the performance experimentally with that of the standard
algorithm.

The computational costs {\em per sweep}, that is,
per $n$ MCMC steps, of the various components of the sparse
algorithm are as follows.
\begin{itemize}
\item Matrix-vector products in GMRES: $2k_1k_2n^2$,
      where $k_1$ is the average number of GMRES iterations per Monte Carlo step
      (per linear system to solve),
      and $k_2 = \mathrm{nnz}(A) / n$ (average number of nonzeros (nnz) in $A$ per row).
\item Computing ILUTP preconditioners:
      $\a k_2 k_3 s n^2$, where $k_3 \ll 1$ is the number of times the preconditioner is
      computed per MCMC step,
      $\a = (\mathrm{nnz}(L) + \mathrm{nnz}(U))/\mathrm{nnz}(A)$,
      and $s$ is the cost per nonzero in the
      preconditioner of computing the ILUTP preconditioner.
      Notice that $\a \leq 2$ by choice (see above), and effectively is
      about $1.25$ (see Table~\ref{table:cost}). The worst case cost of
      computing an ILUTP preconditioner with a constant (small) maximum number of
      nonzeros per row (independent of $n$)
      is $O(n^2)$. However, our timings show that, for this problem, the cost is
      always $O(n)$, which seems to be true in general for the ILUTP preconditioner.
      The parameter $k_3$ should be picked to balance the cost of computing the ILUTP
      preconditioner with the cost of applying the multiplicative updates to the
      preconditioner. Hence, if computing the preconditioner has linear
      cost, $k_3$ should be a constant based on an estimate of the cost of computing
      the ILUTP preconditioner.
\item Applying the preconditioner in GMRES: $2\a k_1k_2 n^2 +
      2 (k_1/k_3) n^2$, where $(1/2)k_3^{-1}$ is the average number of preconditioner
      updates in (\ref{finalMkm}).
\item Matrix reordering: $k_5k_6 n$, where
      $k_5$ is the number of times the reordering is performed per sweep,
      and $k_6$ is the average cost of comparisons and swapping rows (or columns) per row.
      The parameter $k_6$ can vary from nearly constant to $O(n)$; however, it can be brought
      down to a constant by a more elaborate implementation. In general, the cost of the
      reorderings is almost negligible. Moreover, a careful
      incremental implementation should reduce the overall cost
      of reordering to $O(n)$ per sweep.
\end{itemize}
Table~\ref{table:cost} gives a quick overview of experimental
values for the most important parameters. We see that the
average number of GMRES iterations, $k_1$, initially increases
slowly but levels off for larger numbers of particles. The
numbers of nonzeros in the matrix, $k_2$, and in the
preconditioner, $\a k_2$, are roughly constant with $\a \approx
1.25$. Finally, we see that the number of reorderings per
sweep, $k_5$, increases slowly.

\begin{table}[h]
    \begin{center}
        \begin{tabular}{|l|l|l|l|l|l|l|l|l|}
            \hline
            Size & 686 & 1024 & 1458 & 2000 & 2662 & 3456 & 4394 & 5488 \\
            \hline
            nr. GMRES iter.    & 8.91 & 9.34 & 9.53 & 9.59 & 9.83 & 10.20 & 10.22 & 10.10 \\
            nnz(A)/n           & 42.38 & 42.38 & 42.39 & 42.38 & 42.37 & 42.37 & 42.37 & 42.37 \\
            nnz(L+U)/n         & 55.04 & 54.48 & 54.33 & 53.61 & 53.27 & 53.28 & 53.28 & 53.01 \\
            % Upper bound for $m$ & 22 & 34 & 48 & 66 & 88 & 115 & 146 & 182 \\
            nr. reorder/sweep  & 0.65 & 1.03 & 1.19 & 1.73 & 2.23 & 2.63 & 2.73 & 3.12\\
            \hline
        \end{tabular}
    \end{center}
    \caption{Analysis of computational cost, providing for each problem size
    the average number of GMRES iterations per linear system solve ($k_1$), the average number of
    nonzeros per row in the matrix $A$ ($k_2$), the average number of
    nonzeros (per row) in the $L$ and $U$ factors together ($\a k_2$),
    and the average number of matrix reorderings per sweep ($k_5$).}
    \label{table:cost}
\end{table}
Based on the numbers in Table~\ref{table:cost} and the
complexity analysis above, we expect the experimental
complexity of the new algorithm (for the values of $n$ used) to
be slightly worse than $O(n^2)$. However, if the average number
of GMRES iterations ($k_1$) remains bounded for increasing $n$,
as suggested by the table, and we change the reordering
algorithm to a version that is $O(n)$ per reordering, we have
an $O(n^2)$ algorithm.

\begin{figure}[bth]
    \centering
    \includegraphics[scale=0.35]{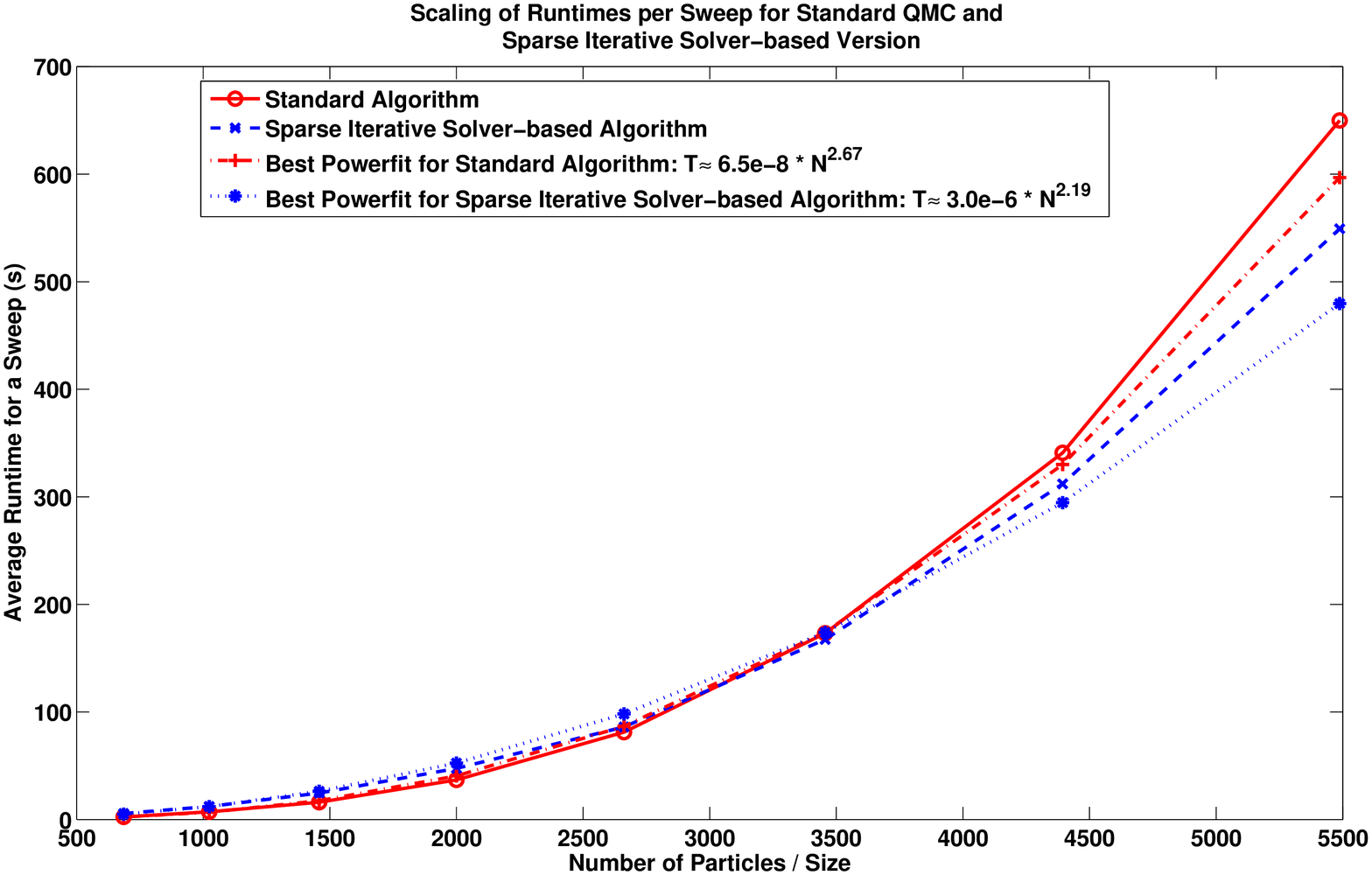}
    \caption{The scaling of the runtime for a sweep (including all computations) for QMCPACK with its
    standard algorithm for determinant ratios and with the new sparse algorithm. The figure also shows
    the best power-law approximations for both versions.}
    \label{figure:QMC_Scaling}
\end{figure}

We compare the average runtimes per sweep of VMC using QMCPACK
with the sparse algorithm with the average runtimes of QMCPACK
with the standard algorithm for eight problems sizes in Table~\ref{table:time}
and Figure~\ref{figure:QMC_Scaling}.
This comparison includes all parts of the
simulation, updating the Slater matrix, computing (some)
observables, and the various parts of computing the determinant
ratios discussed above. For QMCPACK with the sparse algorithm, the
percentage of runtime spent in the linear solver ranges from about $85\%$
for the smallest system to $95\%$ and higher for the larger systems.
For QMCPACK with the standard algorithm the percentages are even higher.
We see that the break-even point for
the new algorithm (for this problem) occurs for about $3000$
particles. Fitting both sequences of runtimes to a power law
($a n^b$), we find that QMCPACK scales as $O(n^{2.67})$ with
the standard algorithm and as $O(n^{2.19})$ with the sparse
algorithm. The exponent for QMCPACK with the sparse algorithm
is slightly larger than $2$, which is partly explained by the
slow increase in the number of GMRES iterations and in the
number of reorderings, although the former appears to level off
for larger $n$. However, we suspect that the exponent is also
partly due to cache and other hardware effects. The exponent
for QMCPACK with the standard algorithm depends on all
components of the algorithm (not just the $O(n^3)$ computation
of determinant ratios); it is noteworthy that it is close to
$3$ already for these system sizes. In the final section, we
discuss how we intend to bring the scaling for the sparse
algorithm down further in future work. We remark that based on
a straightforward complexity analysis, the scaling for the
standard algorithm will approach $O(n^3)$ for large enough $n$.

\begin{table}[bht]
    \begin{center}
        \begin{tabular}{|l|l|l|l|l|l|l|l|l|}
            \hline
            Size & 686 & 1024 & 1458 & 2000 & 2662 & 3456 & 4394 & 5488 \\
            \hline
            std alg. (s) & 2.52 & 7.18 & 16.09 & 36.83 & 81.27 & 173.24 & 340.94 & 649.94 \\
            sparse alg. (s) & 5.71 & 12.18 & 24.67 & 47.55 & 86.11 & 167.43 & 312.05 & 549.17 \\
            \hline
        \end{tabular}
    \end{center}
    \caption{Scaling Results: the average runtime of a VMC sweep using QMCPACK with its standard
    algorithm for determinant ratios and with the new sparse algorithm for determinant ratios.}
    \label{table:time}
\end{table}

\section{Conclusions and Future Work}\label{sec:conclusion}

In this paper, we present an efficient algorithm for simulating
insulators with the VMC method in the limit of large numbers of
particles. Our algorithm reduces the scaling of computing the
determinant ratios, the dominant computational cost for large
$n$, from $O(n^3)$ to slightly worse than $O(n^2)$, where $n$
is the number of particles. This complements recent
improvements in the scaling of constructing Slater matrices.
Our main contribution is a method to compute {\em efficiently}
for the Slater matrices a sequence of preconditioners that make
the iterative solver {\em converge rapidly}. This involves
cheap preconditioner updates, an effective reordering strategy,
and a cheap method to monitor instability of the ILUTP
preconditioners. Furthermore, we demonstrate experimentally
that we can achieve the improved scaling without sacrificing
accuracy. Our results show that preconditioned iterative
solvers can reduce the cost of VMC for large(r) systems.

There are several important improvements to be explored for our
proposed algorithm. First, we will implement an $O(n)$ version
of the reordering algorithm. This is important for larger
system sizes. We will also consider more elaborate reordering
strategies like those in \cite{duff:diag,duff:diag2}. Second,
we intend to develop an incremental local reordering scheme
that satisfies certain optimality properties. This should allow
us to cheaply update the matrix (probably at constant cost)
every Monte Carlo step or every few steps and exploit the fact
that particle updates are strictly local. Potentially, the
field of computational geometry might provide some insights for
this effort. Third, although the ILUTP preconditioner leads to
fast convergence, it is not obvious how to update the ILUTP
preconditioner in its LU form. Again, the current approach with
cheap intermediate, multiplicative updates by rank-one updates
to the identity is effective, but it would be better to have
preconditioners that can be updated continually with constant
cost, local updates, and that adapt in a straightforward manner
to a reordering of the matrix (at constant cost). We do not
know if such preconditioners exist and whether they would yield
fast convergence. In future work, we will explore forms of
preconditioning that might satisfy these requirements and an
underlying theory of preconditioners for Slater matrices. The
latter would also include analyzing the matrix properties of
Slater matrices. Fourth, an interesting experiment (also suggested
by the referees) is to check whether replacing the (sparse) iterative
solver by a sparse direct solver might be advantageous for certain
problem sizes. We expect that for small problems the standard algorithm
is fastest and for large(r) problems sparse iterative solvers are
fastest. However, there might be a range of physically relevant problem
sizes for which sparse direct solvers are the best.
Fifth, we will extend our algorithm to other
types of orbitals. One specific goal will be to adapt our
algorithm to achieve quadratic scaling when the single particle
orbitals are delocalized. Note that the optimization of
orbitals \cite{dario:school,drummond:school,williamson:linear}
discussed in the Introduction leads to decaying orbitals for
many systems. Finally, we will test our algorithm for realistic
materials and much larger system sizes.

%\bibliographystyle{abbrv}
%\bibliography{vmc}

\bibliographystyle{abbrv}
\bibliography{vmc_upd,cmg}

\begin{thebibliography}{10}

\bibitem{dario:school}
D.~Alf{\' e}.
\newblock Order({N}) methods in {QMC}.
\newblock 2007 Summer School on Computational Materials Science,
  http://www.mcc.uiuc.edu/summerschool/2007/qmc/.

\bibitem{dario:splines}
D.~Alf\'{e} and M.~J. Gillan.
\newblock An efficient localized basis set for {Q}uantum {M}onte {C}arlo
  calculations on condensed matter.
\newblock {\em Physical Review B (Rapids)}, 70:161101 (1--4), 2004.

\bibitem{dario:linear}
D.~Alf\'{e} and M.~J. Gillan.
\newblock Linear-scaling {Q}uantum {M}onte {C}arlo with non-orthogonal
  localized orbitals.
\newblock {\em Journal of Physics: Condensed Matter}, 16:L305--L311, 2004.

\bibitem{AllTildes87}
M.~P. Allen and D.~J. Tildesley.
\newblock {\em Computer Simulation of Liquids}.
\newblock Oxford University Press, 1987.

\bibitem{BaiSca09}
Z.~Bai, W.~Chen, R.~Scalettar, and I.~Yamazaki.
\newblock Numerical methods for {Q}uantum {M}onte {C}arlo simulations of the
  {H}ubbard model.
\newblock In T.~Y. Hou, C.~Liu, and J.-G. Liu, editors, {\em Multi-Scale
  Phenomena in Complex Fluids, ISBN-9787040173581}. Higher Education Press,
  China, February 2009.
\newblock An early version appeared as Technical Report CSE-2007-36, Department
  of Computer Science, UC Davis, Dec.4, 2007 and revised on Feb.25, 2008.

\bibitem{benzi:survey}
M.~Benzi.
\newblock Preconditioning techniques for large linear systems: A survey.
\newblock {\em Journal of Computational Physics}, 182(2):418 -- 477, 2002.

\bibitem{benzi:orderings}
M.~Benzi, D.~B. Szyld, and A.~van Duin.
\newblock Orderings for incomplete factorization preconditioning of
  nonsymmetric problems.
\newblock {\em SIAM Journal on Scientific Computing}, 20(5):1652--1670, 1999.

\bibitem{BerBru06}
L.~Bergamaschi, R.~Bru, A.~Martinez, and M.~Putti.
\newblock Quasi-{N}ewton preconditioners for the inexact {N}ewton method.
\newblock {\em Electronic Transaction on Numerical Analysis}, 23:76--87, 2006.
\newblock http://etna.mcs.kent.edu/vol.23.2006/pp76-87.dir/pp76-87.pdf.

\bibitem{BoyGill11}
J.~P. Boyd and K.~W. Gildersleeve.
\newblock Numerical experiments on the condition number of the interpolation
  matrices for radial basis functions.
\newblock {\em Applied Numerical Mathematics}, 61:443--459, 2011.

\bibitem{CalSom07}
D.~Calvetti and E.~Somersalo.
\newblock {\em Introduction to Bayesian Scientific Computing: Ten Lectures on
  Subjective Computing}, volume~2 of {\em Surveys and Tutorials in the Applied
  Mathematical Sciences}.
\newblock Springer, 2007.

\bibitem{ceperley:tool}
D.~M. Ceperley.
\newblock Data{S}pork analysis toolkit.
\newblock Materials Computation Center at the University of Illinois at
  Urbana-Champaign, http://www.mcc.uiuc.edu/dataspork/.

\bibitem{CepChes77}
D.~M. Ceperley, G.~V. Chester, and M.~H. Kalos.
\newblock {M}onte {C}arlo simulation of a many-fermion system.
\newblock {\em Phys. Rev. B}, 3081(16):3081–--3099, 1977.

\bibitem{chernesky:physics}
M.~P. Chernesky.
\newblock On preconditioned {K}rylov subspace methods for discrete
  convection-diffusion problems.
\newblock {\em Numerical Methods for Partial Differential Equations},
  13(4):321--330, 1997.

\bibitem{chow:vector}
E.~Chow and Y.~Saad.
\newblock Experimental study of {ILU} preconditioners for indefinite matrices.
\newblock {\em Journal of Computational and Applied Mathematics}, 86(2):387 --
  414, 1997.

\bibitem{drummond:school}
N.~Drummond and P.~L. Rios.
\newblock Worksheet 1: Using localized orbitals in {QMC} calculations.
\newblock 2007 Summer School on Computational Materials Science,
  http://www.mcc.uiuc.edu/summerschool/2007/qmc/.

\bibitem{duff:diag}
I.~S. Duff and J.~Koster.
\newblock The design and use of algorithms for permuting large entries to the
  diagonal of sparse matrices.
\newblock {\em SIAM Journal on Matrix Analysis and Applications},
  20(4):889--901, 1999.

\bibitem{duff:diag2}
I.~S. Duff and J.~Koster.
\newblock On algorithms for permuting large entries to the diagonal of a sparse
  matrix.
\newblock {\em SIAM Journal on Matrix Analysis and Applications},
  22(4):973--996, 2001.

\bibitem{edmunds:io}
J.~M. Edmunds.
\newblock Input and output scaling and reordering for diagonal dominance and
  block diagonal dominance.
\newblock {\em IEE Proceedings - Control Theory and Applications},
  145(6):523--530, 1998.

\bibitem{foulkes:review}
W.~M.~C. Foulkes, L.~Mitas, R.~J. Needs, and G.~Rajagopal.
\newblock Quantum {M}onte {C}arlo simulations of solids.
\newblock {\em Rev. Mod. Phys.}, 73(1):33--83, 2001.

\bibitem{FreSmi02}
D.~Frenkel and B.~Smit.
\newblock {\em Understanding Molecular Simulation, Second Edition: From
  Algorithms to Applications}, volume~1 of {\em Computational Science Series}.
\newblock Academic Press, 2002.

\bibitem{golub:book}
G.~H. Golub and C.~F. {Van Loan}.
\newblock {\em Matrix Computations}.
\newblock The Johns Hopkins University Press, 1996.

\bibitem{GreeBk97}
A.~Greenbaum.
\newblock {\em Iterative Methods for Solving Linear Systems}.
\newblock SIAM, 1997.

\bibitem{Ha94}
B.~Hammond, W.~A. Lester, and P.~J. Reynolds.
\newblock {\em {M}onte {C}arlo Methods in Ab Initio Quantum Chemistry}.
\newblock World Scientific, Singapore, 1994.

\bibitem{kim:qmc}
{J. Kim et al.}
\newblock {QMCPACK}.
\newblock Materials Computation Center at the University of Illinois at
  Urbana-Champaign, http://cms.mcc.uiuc.edu/qmcpack/.

\bibitem{kelley:book}
C.~T. Kelley.
\newblock {\em Solving Nonlinear Equations with Newton's Method}.
\newblock Society for Industrial and Applied Mathematics, 2003.

\bibitem{McMi65}
W.~McMillan.
\newblock Ground state of liquid helium$^4$.
\newblock {\em Physical Review}, 138(2A):A442 -- A451, April 1965.

\bibitem{mitas:few}
L.~Mitas and J.~C. Grossman.
\newblock Quantum {M}onte {C}arlo study of {S}i and {C} molecular systems.
\newblock {\em in {R}ecent {A}dvances in {Q}uantum {M}onte {C}arlo {M}ethods,
  W. A. Lester (Ed.)}, pages 133--162, 1997.

\bibitem{nukula:slater}
P.~K. V.~V. Nukala and P.~R.~C. Kent.
\newblock A fast and efficient algorithm for {S}later determinant updates in
  {Q}uantum {M}onte {C}arlo simulations.
\newblock {\em J. Chem. Phys.}, 130(204105), 2009.

\bibitem{Saad94}
Y.~Saad.
\newblock {ILUT}: a dual threshold incomplete {ILU} factorization.
\newblock {\em Numer. Linear Algebra Appl.}, 1:387--402, 1994.

\bibitem{SaadBk03}
Y.~Saad.
\newblock {\em Iterative Methods for Sparse Linear Systems}.
\newblock Society for Industrial and Applied Mathematics, 3600 Market Street,
  Philadelphia, PA 19104-2688, USA, 2nd edition, 2003.

\bibitem{SaadSchu86}
Y.~Saad and M.~H. Schultz.
\newblock {GMRES}: A generalized minimal residual algorithm for solving
  nonsymmetric linear systems.
\newblock {\em SIAM Journal on Scientific and Statistical Computing},
  7(3):856--869, 1986.

\bibitem{scalettar:few}
R.~T. Scalettar, K.~J. Runge, J.~Correa, P.~Lee, V.~Oklobdzija, and J.~L.
  Vujic.
\newblock Simulations of interacting many body systems using p4.
\newblock {\em International Journal of High Speed Computing}, 7(3):327--349,
  1995.

\bibitem{thijssen:book}
J.~Thijssen.
\newblock {\em Computational Physics}.
\newblock Cambridge University Press, 1999.

\bibitem{VorsBk03}
H.~A. van~de Vorst.
\newblock {\em Iterative Krylov Methods for Large Linear Systems}.
\newblock Cambridge Monographs on Applied and Computational Mathematics.
  Cambridge University Press, 2003.

\bibitem{williamson:linear}
A.~J. Williamson, R.~Q. Hood, and J.~C. Grossman.
\newblock Linear-scaling {Q}uantum {M}onte {C}arlo calculations.
\newblock {\em Phys. Rev. Lett.}, 87(24):246406, 2001.

\end{thebibliography}

\end{document}